\documentclass[fleqn,usenatbib]{mnras}

\usepackage{newtxtext,newtxmath}
\usepackage{lineno}
\usepackage[T1]{fontenc}

\DeclareRobustCommand{\VAN}[3]{#2}
\let\VANthebibliography\thebibliography
\def\thebibliography{\DeclareRobustCommand{\VAN}[3]{##3}\VANthebibliography}

\usepackage{graphicx}	% Including figure files
\usepackage{amsmath}	% Advanced maths commands
\usepackage{amssymb}	% Extra maths symbols

\usepackage{natbib}

\makeatletter

\patchcmd{\NAT@citex}
  {\@citea\NAT@hyper@{%
     \NAT@nmfmt{\NAT@nm}%
     \hyper@natlinkbreak{\NAT@aysep\NAT@spacechar}{\@citeb\@extra@b@citeb}%
     \NAT@date}}
  {\@citea\NAT@nmfmt{\NAT@nm}%
   \NAT@aysep\NAT@spacechar\NAT@hyper@{\NAT@date}}{}{}

\patchcmd{\NAT@citex}
  {\@citea\NAT@hyper@{%
     \NAT@nmfmt{\NAT@nm}%
     \hyper@natlinkbreak{\NAT@spacechar\NAT@@open\if*#1*\else#1\NAT@spacechar\fi}%
       {\@citeb\@extra@b@citeb}%
     \NAT@date}}
  {\@citea\NAT@nmfmt{\NAT@nm}%
   \NAT@spacechar\NAT@@open\if*#1*\else#1\NAT@spacechar\fi\NAT@hyper@{\NAT@date}}
  {}{}

\makeatother

\usepackage[dvipsnames]{xcolor}
\hypersetup{
    colorlinks=true,
    linkcolor=blue,      
    urlcolor=DarkOrchid,
    citecolor=blue
}

\title["Life" of dust from Jupiter's irregular satellites]{"Life" of dust originating from the irregular satellites of Jupiter}
\author[Zhenghan Chen et al.]{
Zhenghan Chen,$^{1,2}$
Kun Yang,$^{1,2}$
and Xiaodong Liu$^{1,2}$\thanks{E-mail: liuxd36@mail.sysu.edu.cn (XL)}
\\
$^{1}$School of Aeronautics and Astronautics, Shenzhen Campus of Sun Yat-sen University, Shenzhen, Guangdong 518107, China\\
$^{2}$Shenzhen Key Laboratory of Intelligent Microsatellite Constellation, Shenzhen Campus of Sun Yat-sen University, Shenzhen, Guangdong 518107, China\\
}

\date{Accepted XXX. Received YYY; in original form ZZZ}

\pubyear{2023}

\begin{document}
\label{firstpage}
\pagerange{\pageref{firstpage}--\pageref{lastpage}}
\maketitle

\begin{abstract}
%This is a simple template for authors to write new MNRAS papers.
%The abstract should briefly describe the aims, methods, and main results of %the paper.
%It should be a single paragraph not more than 250 words (200 words for %Letters).
%No references should appear in the abstract.
The irregular satellites of Jupiter produce dust particles through the impact of interplanetary micrometeoroids.
In this paper, the dynamics of these particles is studied by both high-accuracy numerical simulation and analytical theory, in order to learn their transport, final fate and spatial distribution.
The perturbation forces that are considered in our dynamical model include the solar radiation pressure, solar gravity, Poynting-Robertson drag, Jovian oblateness and the Galilean satellites’ gravity. The trajectories of different size particles are simulated until they hit Jupiter, the Galilean satellites or escape from the Jovian system.
The average dynamical lifetimes of dust with different grain sizes are calculated, and the final fate of dust particles is reported and analyzed. The steady-state spatial number density of particles is estimated by integrating the trajectories of dust particles over their initial size distribution, and compared to the previous work. The impact sites of dust on Callisto's surface are recorded and provide an important clue for the study of the hemisphere asymmetry of Callisto. Besides, the mass accretion rate, cross-sectional area influx and mass influx density of dust on Callisto are calculated.
A ring outside the orbit of Callisto dominated by dust between 2 and 25 $\mu$m from Jupiter's irregular satellites is suggested,
%with the normal optical depth on the same order with the Phoebe ring and the configuration of ring ansae similar to Jupiter’s gossamer rings. 
with the average normal geometric optical depth on the order of $10^{-8}$ and the configuration of the ring ansae similar to Jupiter’s gossamer rings.
\end{abstract}


%%%%%%%%%%%%%%%%%%%%%%%%%%%%%%%%%%%%%%%%%%%%%%%%%%

%%%%%%%%%%%%%%%%% BODY OF PAPER %%%%%%%%%%%%%%%%%%

\section{Introduction}
%This is a simple template for authors to write new MNRAS papers.
%See \texttt{mnras\_sample.tex} for a more complex example, and \texttt{mnras\_guide.tex}
%for a full user guide.

%All papers should start with an Introduction section, which sets the work
%in context, cites relevant earlier studies in the field by \citet{Fournier1901},
%and describes the problem the authors aim to solve \citep[e.g.][]{vanDijk1902}.
%Multiple citations can be joined in a simple way like \citet{deLaguarde1903, delaGuarde1904}.

\begin{figure}
	\includegraphics[width=\columnwidth]{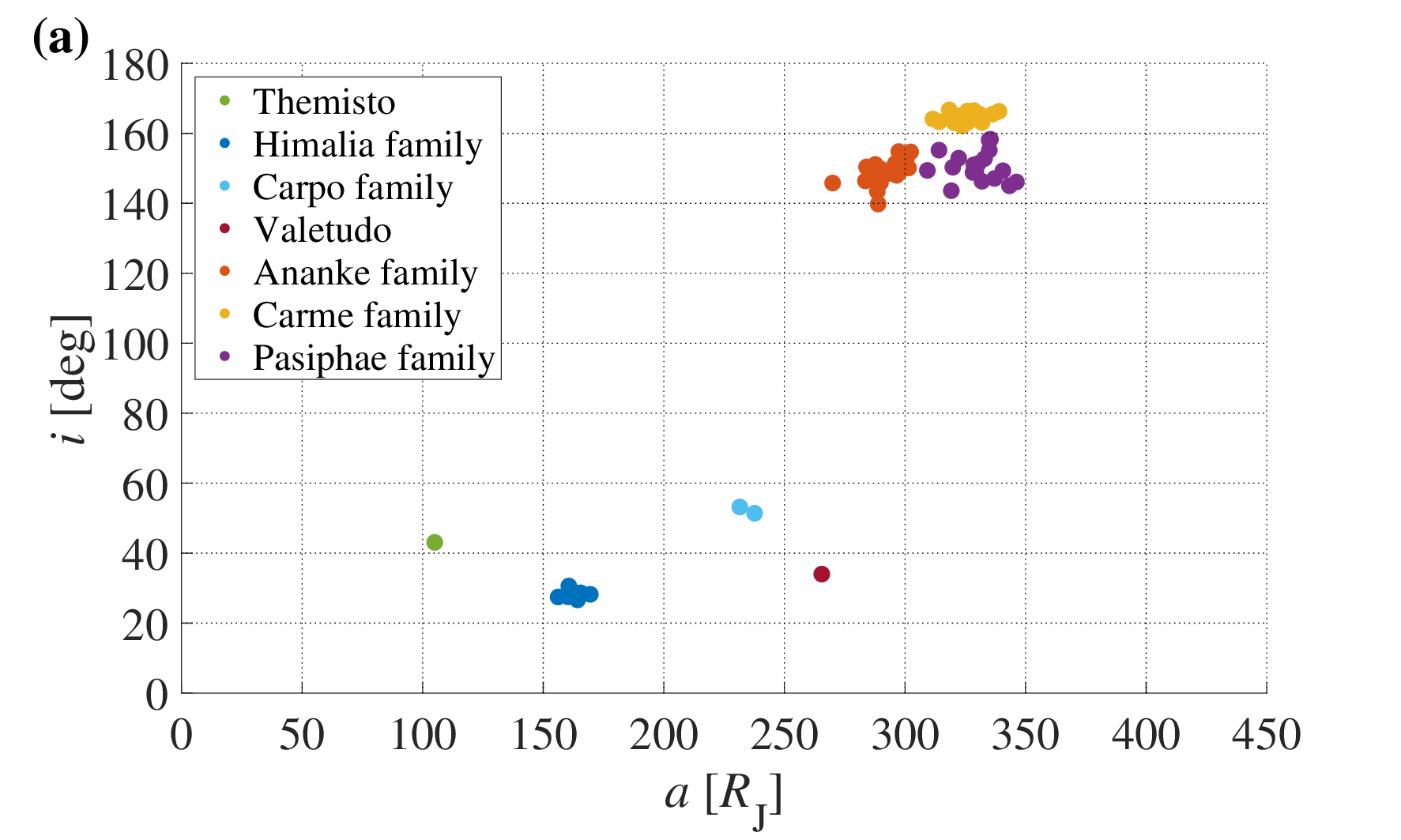}
        \includegraphics[width=\columnwidth]{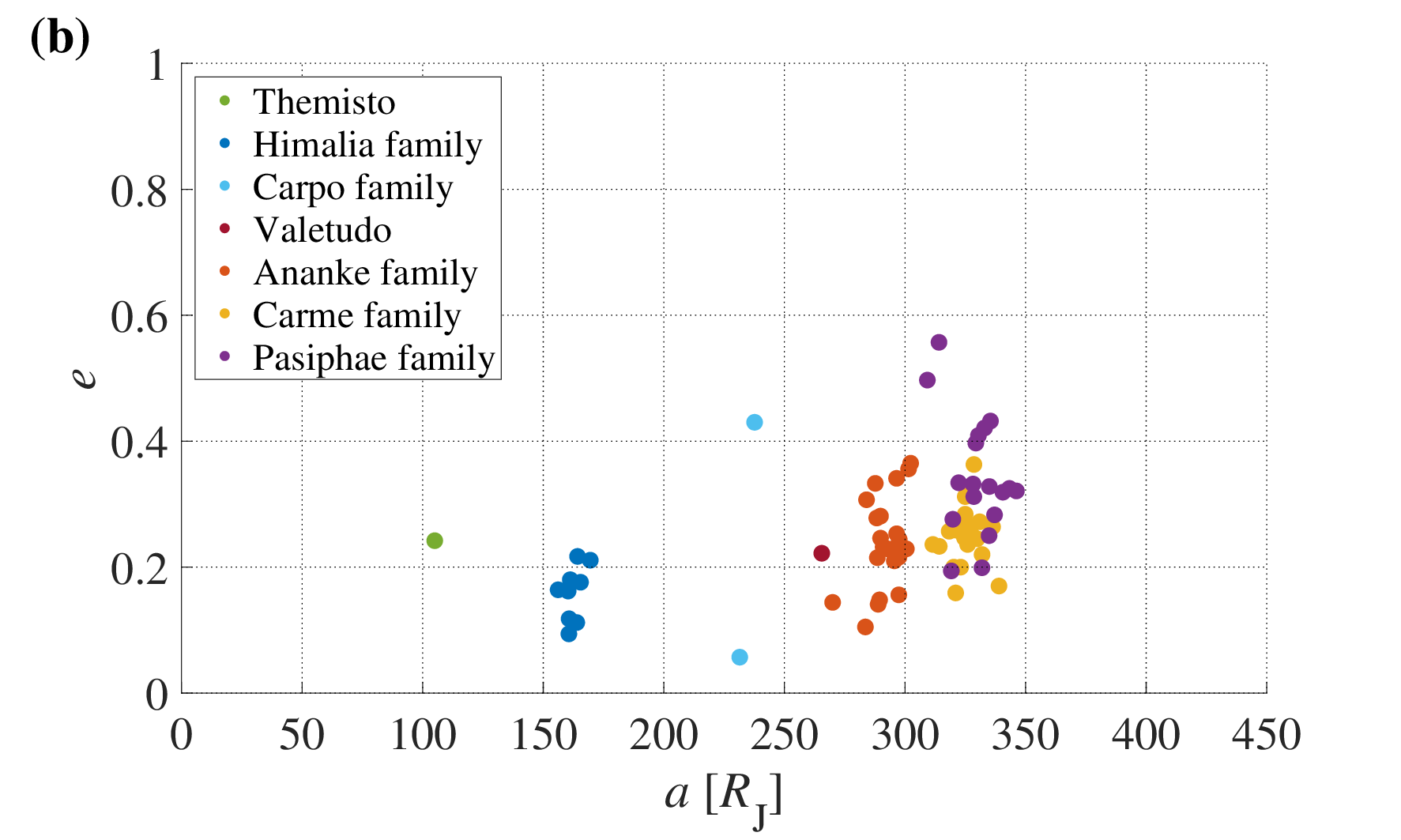}
        \includegraphics[width=\columnwidth]{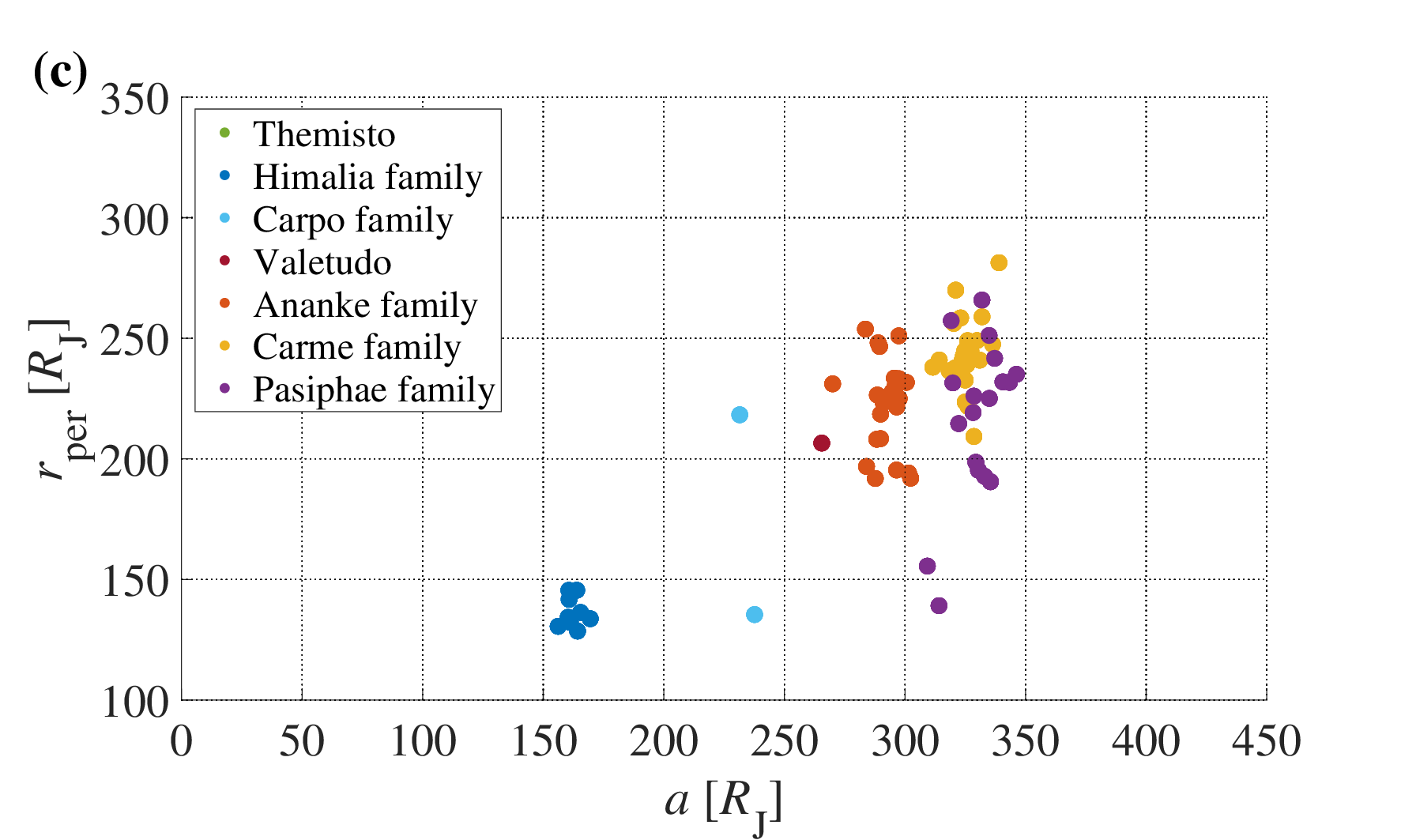}
    \caption{(a) The semimajor axis vs.~orbital inclination, for irregular satellites of Jupiter. The data is obtained from \url{https://sites.google.com/carnegiescience.edu/sheppard/moons/jupitermoons}.
    (b) The semimajor axis vs.~orbital eccentricity for the same moons in (a).
    (c) The semimajor axis vs.~distance of orbital pericenter relative to Jupiter for the same moons in (a).}
    \label{fig1}
\end{figure}

Jupiter has a lot of small irregular satellites located far from the central planet and generally with large inclinations and eccentricities (see Fig.~\ref{fig1}). As other airless celestial bodies in the universe, these moons are constantly bombarded by high-speed interplanetary micrometeoroids to generate a large amount of dust, which has complicated dynamical evolution under the effect of various perturbation forces. Data obtained by the Galileo dust detector system (DDS) indicates that dust originating from the irregular satellites travels in the outer region of Jupiter \cite[50-300 Jovian radii,][]{krivov2002dust}, and the recent reanalysis of the DDS data suggests that particles from the irregular moons may also contribute to the dust in the territories of Galilean satellites \cite[]{soja2015new}.

However, there are not many studies on the dynamics of dust ejected from Jupiter’s irregular satellites. To compare with the DDS data, \cite{krivov2002dust} analyzed the motion of dust particles by a simplified analytical model with numerical simulations over less than a million years, which mainly considered the influence induced by solar radiation pressure and solar gravity.
%Although a more detailed dynamical equation and the long-term integration were applied by \cite{bottke2013black} to explore the accretion of irregular satellite debris onto the Galilean moons, less attention were paid to the fates of small dust particles and their spatial distribution.
The long-term numerical integration of irregular satellite debris was applied by \cite{bottke2013black} to explore their accretion on the Galilean moons. Less attention was paid to the fate of small dust particles and their spatial distribution.
Besides, the recent observed “Phoebe” ring \cite[]{verbiscer2009saturn}, which is thought to be formed by the dust released from the Saturn’s irregular moon Phoebe, enlightens us to seek whether Jupiter has a similar ring structure.

In this paper, the dynamical evolution of dust particles with 12 different grain sizes between 1 and 500 $\mu$m originating from four groups of Jupiter's irregular satellites is simulated with high-accuracy numerical integrations. The average dynamical lifetimes of different size particles are calculated and analyzed theoretically.
%The final fate and the impact site on Callisto's surface of dust from different satellite families are investigated by analyzing the distribution of orbital elements of particles after they arrive at the regions of Galilean satellites.
The final fate of dust from different satellite families is investigated by analyzing the distribution of orbital elements of dust. The trajectories of dust grains with different sizes are integrated over their initial size distribution to estimate the steady-state number density of dust in the Jovian system. The impact sites of dust particles on Callisto's surface are recorded, and the mass accretion rate, cross-sectional area influx and mass influx density of dust on Callisto are calculated. The normal and edge-on geometric optical depths of the ring outside the orbit of Callisto are estimated, and the contributions of dust from different satellite families to the ring are evaluated.

This paper is organized as follows.
The production rate and initial size distribution of dust from Jupiter's irregular satellites are derived in Section 2. The dynamical model and simulation procedure are described in Section 3. In Section 4, the numerical simulation results, including the average dynamical lifetimes, the final fate of dust, the spatial number density, the impact sites and the accretion of dust on Callisto's surface, and the geometric optical depth, are presented and analyzed.
Finally, the summary of our work is given in Section 5.

\section{Depart from irregular satellites}
%Normally the next section describes the techniques the authors used.
%It is frequently split into subsections, such as Section~\ref{sec:maths} below.

\subsection{Production rate}
\label{sec:product} % used for referring to this section from elsewhere
The impact-ejection process, one of the major source of the circumplanetary dust, has been well-studied.
We use the formula derived by \cite{krivov2003impact} to estimate the production rate of dust ejected from the surfaces of Jupiter’s irregular satellites, which reads
\begin{equation}
    M^{+}=F_\mathrm{imp}YS.
	\label{eq:1}
\end{equation}
Here $F_\mathrm{imp}$ is the mass flux of the impactor, $Y$ is the ejecta yield which depends on the composition of the target's surface and projectile’s speed and mass, and $S$ is the cross-sectional area of the target bodies suffered to impact.

%\textcolor{red}{The mass flux and velocity of the interplanetary micrometeoroids within the Jovian system can be estimated by correcting the values before they enter the Jovian system for the effect of the gravitational focusing by Jupiter, which reads \cite[]{spahn2006ring}}
When taking into account the effect of the gravitational focusing by Jupiter, the mass flux of the interplanetary micrometeoroids and its corresponding average velocity in the Jupiter system read \cite[]{spahn2006ring}
\begin{equation}
\label{eq: Fimp}
\resizebox{\columnwidth}{!}{$
\begin{split}
    F_\mathrm{imp} &= \frac{1}{2}F_\mathrm{imp}^{\infty}\frac{V_\mathrm{imp}}{V_\mathrm{imp}^{\infty}}\\
    &\left\{\sqrt{1+\frac{2GM_\mathrm{J}}{a_\mathrm{t}\left(V_\mathrm{imp}^{\infty}\right)^2}}+\sqrt{1+\frac{2GM_\mathrm{J}}{a_\mathrm{t}\left(V_\mathrm{imp}^{\infty}\right)^2}-\left(\frac{R_\mathrm{J}}{a_\mathrm{t}}\right)^2\left[1+\frac{2GM_\mathrm{J}}{R_\mathrm{J}\left(V_\mathrm{imp}^{\infty}\right)^2}\right]}\right\},
    \end{split}
     $}
\end{equation}
\begin{equation}
\label{eq: Vimp}
    V_\mathrm{imp} = V_\mathrm{imp}^{\infty}\sqrt{1+\frac{2GM_\mathrm{J}}{a_\mathrm{t}\left(V_\mathrm{imp}^{\infty}\right)^2}}.
\end{equation}
Here $a_\mathrm{t}$ is the distance of the target moon relative to Jupiter, which is taken as the moon's semimajor axis approximately, $F_\mathrm{imp}^{\infty}=10^{-16}\,\mathrm{kg}\,\mathrm{m^{-2}}\,\mathrm{s^{-1}}$ and $V_\mathrm{imp}^{\infty}=3.7\,\mathrm{km}\,\mathrm{s^{-1}}$ are the mass flux and the average velocity of the interplanetary dust particles at the heliocentric distance of Jupiter without considering the gravitational focusing by Jupiter, respectively, the values of which are adopted from \cite{poppe2016improved}. The variable $G$ is the gravitational constant, and $M_\mathrm{J}$ is the mass of Jupiter.
It is obvious from Equations (\ref{eq: Fimp}-\ref{eq: Vimp}) that the effect of the gravitational focusing by Jupiter is stronger at a closer distance to Jupiter.
%The values of $F_\mathrm{imp}$ and $V_\mathrm{imp}$ when considering the gravitational focusing effect by Jupiter at the distances of four families of Jupiter's irregular satellites are listed in Table \ref{tab: production rate}.

The ejecta yield $Y$ is calculated by the empirical formula \cite[]{koschny2001impacts, koschny2001impacts-b}
\begin{equation}
\resizebox{\columnwidth}{!}{$
\begin{split}
    Y=2.85\,{\times}\,10^{-8}\,{\times}\,0.0149^{G_\mathrm{sil}}\,{\times}\,\left(\frac{1-G_\mathrm{sil}}{\rho_\mathrm{ice}}+\frac{G_\mathrm{sil}}{\rho_\mathrm{sil}}\right)^{-1}m_\mathrm{imp}^{0.23}v_\mathrm{imp}^{2.46}.
 \end{split}
 $}
\end{equation}
%\textcolor{red}{which assumes that the composition of the target's surface is a linear mixture of ice and silicate, with the density $\rho_\mathrm{ice}=927\,\mathrm{kg/m^3}$ for ice and $\rho_\mathrm{sil}=2800\,\mathrm{kg/m^3}$ for silicate. Therefore, the fraction of silicate of the surface $G_\mathrm{sil}$ can be estimated through the density of the moon $\rho_\mathrm{t}$, i.e.}
Here the surfaces of the satellites are assumed to be mixtures of silicate and ice, $\rho_\mathrm{sil}=2800\,\mathrm{kg\,m^{-3}}$ and $\rho_\mathrm{ice}=927\,\mathrm{kg\,m^{-3}}$ are the densities of silicate and ice, respectively, and $m_\mathrm{imp}=10^{-8}\,\mathrm{kg}$ is the typical mass of the impactor \citep{krivov2003impact}. The variable $G_\mathrm{sil}$ is the fraction of silicate in the satellite's surface material, the value of which can be estimated by the formula \cite[]{koschny2001impacts, koschny2001impacts-b}
\begin{equation}
\begin{split}
    \rho_\mathrm{t} = \left(\frac{1-G_\mathrm{sil}}{\rho_\mathrm{ice}}+\frac{G_\mathrm{sil}}{\rho_\mathrm{sil}}\right)^{-1},
 \end{split}
\end{equation}
where $\rho_\mathrm{t}$ is the density of the target moon.

% The production rates of four main families, which \textcolor{red}{are} named of their largest moon, Himalia, Carme, Pasiphae and Ananke, are calculated rather than of each satellite because of the clustering properties of Jupiter’s irregular moons \cite[see Fig.~\ref{fig1};][]{jewitt2007irregular}, with $S$ of each family equal to the sum of their moon member’s cross-sectional area. The result is listed in Table \ref{tab1}.
The values of $a_\mathrm{t}$, $\rho_\mathrm{t}$, $F_\mathrm{imp}$, $V_\mathrm{imp}$, $S$, $Y$, and $M^+$ of four main families, which are named of their respective largest moons, i.e.~Himalia, Ananke, Carme and Pasiphae, are calculated and listed in Table \ref{tab: production rate}. It is seen that the mass production rate of dust from the prograde Himalia family is much greater than those of the three retrograde families, which is due to the stronger gravitational focusing effect by Jupiter on interplanetary particles in the Himalia region and the larger $S$ of the Himalia group.

\begin{table*}
        \renewcommand{\arraystretch}{1.5}
        \caption{Assumed values of quantities associated with the four satellite families, including semimajor axis ($a_\mathrm{t}$), bulk density ($\rho_\mathrm{t}$), approximate cross-sectional area ($S$) of the satellite family suffered to impact, velocity ($V_\mathrm{imp}$) and mass flux ($F_\mathrm{imp}$) of impactors when considering the gravitational focusing effect, ejecta yield ($Y$) and mass production rate ($M^+$). Note that the values of $a_\mathrm{t}$ and $\rho_\mathrm{t}$ of the four satellite families are taken as those of their respective largest satellites. The density of Himalia is calculated according to the estimations of its mass of $4.2{\times}10^{18}\,\mathrm{kg}$ \citep{emelyanov2005mass} and average radius of $85\,\mathrm{km}$ \citep{archinal2018report}. The bulk densities of other three satellite families are assumed to be the same as the typical density values of asteroids with the same spectral types, where the densities of Ananke, Carme, and Pasiphae are taken as those of P-type \citep{grav2004near,vernazza2021vlt}, D-type \citep{grav2015neowise, carry2012density} and C-type \citep{grav2004near,vernazza2021vlt} asteroids, respectively.}
        \label{tab: production rate}
	\begin{tabular*}{\textwidth}{@{\extracolsep{\fill}}ccccccccc} % four columns, alignment for each
		\hline
		Satellite family & Type of orbit & $a_\mathrm{t}\,[R_\mathrm{J}]$ & $\rho_\mathrm{t}\,[\mathrm{g\,m^{-3}}]$ & $S\,\mathrm{[m^2]}$ & $V_\mathrm{imp}\,[\mathrm{km\,s^{-1}}]$ & $F_\mathrm{imp}\,[\mathrm{kg}\,\mathrm{m^{-2}}\,\mathrm{s^{-1}}]$ & $Y$ & $M^+\,[\mathrm{kg\,s^{-1}}]$\\
		\hline
		Himalia & prograde & 160.3 & 1.63 & 7.55E10 & 5.98 & 2.61E-16 & 8.72E1 & 1.70E-3 \\
            Ananke & retrograde & 297.6 & 1.30 & 6.22E9 & 5.06 & 1.87E-16 & 1.14E2 & 1.33E-4 \\
		Carme & retrograde & 327.4 & 1.63 & 9.85E9 & 4.95 & 1.79E-16 & 5.47E1 & 9.65E-5 \\    
		Pasiphae & retrograde & 330.4 & 1.70 & 1.56E10 & 4.94 & 1.78E-16 & 4.90E1 & 1.36E-4 \\
		\hline
	\end{tabular*}
\end{table*}

\subsection{Initial size distribution}
\label{sec:Initial size distribution}
The initial size distribution of ejected particles is assumed to follow a power law,
\begin{equation}
    N(r_\mathrm{g}) \, \mathrm{d}r_\mathrm{g} = C_0 \, r_\mathrm{g}^{-q} \, \mathrm{d}r_\mathrm{g},
        \label{eq:2}
\end{equation}
where $N(r_\mathrm{g})\mathrm{d}r_\mathrm{g}$ is the number of dust particles that are ejected per unit time in the size range of [$r_\mathrm{g}$, $r_\mathrm{g}+\mathrm{d}r_\mathrm{g}$], $C_0$ is the normalization constant, and $q=3.73$ is the exponent adopted from \cite{horanyi2015permanent}, which was obtained from the in-situ measurements in the lunar dust environment. By normalizing the distribution to the production rate
\begin{equation}
    \int_{r_\mathrm{min}}^{r_\mathrm{max}}\frac{4}{3}{\pi} \, r_\mathrm{g}^{3} \,{\rho} \, N(r_\mathrm{g}) \, \mathrm{d}r_\mathrm{g}=M^+,
        \label{eq:3}
\end{equation}
we can get
\begin{equation}
    N(r_\mathrm{g})=\frac{3M^+}{4{\pi}{\rho}}\frac{4-q}{r_\mathrm{max}^{4-q}-r_\mathrm{min}^{4-q}}r_\mathrm{g}^{-q}.
        \label{eq:4}
\end{equation}
Here $r_\mathrm{max}$, $r_\mathrm{min}$ are the maximal and minimal radii of ejected particles, and $\rho$ is the dust’s bulk density which is assumed to be the same as that of the parent moon (see Table \ref{tab: production rate}).

% Due to the small size of irregular satellites, almost all of the ejecta can escape from their parent moon’s gravity with the speed close to the escape velocity \cite[]{szalay2018dust}.

\section{dynamical simulation}
\subsection{The equation of dust's motion}
\label{sec:The equation of dust's motion}
The escape velocities of Jupiter's irregular satellites are very low due to their small sizes. Even for the largest irregular moon Himalia, the escape velocity is only about $0.06\,\mathrm{km\,s^{-1}}$. Therefore, almost all of the ejecta can escape from Jupiter's irregular satellites.

After leaving the parent bodies, dust is subject to various forces.
%Solar radiation pressure and solar gravity are dominant perturbation in the irregular satellites region \cite[]{hamilton1996circumplanetary}, which cause the orbital eccentricity of dust to oscillate.
Solar radiation pressure is important for small particles \cite[]{burns1979radiation}, and solar gravity is also a significant perturbation since the irregular satellites are far from the Jupiter \cite[]{hamilton1996circumplanetary}. Poynting-Robertson drag (PR drag) leads dust to migrate toward the central planet slowly \cite[]{burns1979radiation}.
%The Jovian oblateness $J_2$ and the regular moons’ gravity are also needed to be considered, as the dust can encounter with Jupiter and Galilean satellites by getting into high eccentric orbit, or by inward migration.
The Jovian oblateness and gravitational perturbations from the outer Galilean satellites also need to be considered.
%as the dust can encounter with the Jupiter and Galilean moons when the orbital eccentricity is large, or after a long period of inward migration. 
In this work, the motion of dust particles for 12 grain sizes in the range of $[1, 500]\,{\mu}$m is simulated in the Jovian equatorial inertial frame (JIF). The $x$-axis of JIF points to the ascending node of Jupiter's orbit, the $z$-axis is along the spin axis of Jupiter, and the $y$-axis is defined by right-hand rule. The equation of dust particle's motion in JIF is expressed as
 %\ddot{\vec{r}}=-\frac{GM_\mathrm{J}}{r^3}\Vec{r}+(\frac{SA}{c})Q_\mathrm{pr}(1-\frac{\Vec{v}\cdot\Vec{S}}{c})\Vec{S}-[\frac{GM_{sun}}{{|\Vec{r}-\Vec{r}_{sun}|}^3}+\frac{GM_{sun}}{|\Vec{r}_{sun}|^3}\Vec{r}_{sun}]-(\frac{SA}{c})Q_{pr}\frac{\Vec{v}}{c}+GM_{J}R_{J}^{2}J_{2}\triangledown(\frac{P_{2}(\Vec{S}\cdot\Vec{r})}{r_{3}})-[\frac{GM_{Galilean}}{{|\Vec{r}-\Vec{r}_{Galilean}|}^3}+\frac{GM_{Galilean}{|\Vec{r}_{Galilean}|^3}\Vec{r}_{Galilean}],
\begin{equation}
    \begin{split}
\ddot{\vec{r}}&=-\frac{GM_\mathrm{J}}{r^3}\Vec{r}+\left(\frac{S_\mathrm{flux}A_\mathrm{g}}{m_\mathrm{g}c}\right)Q_\mathrm{pr}\left[1-\frac{(\Dot{\Vec{r}}-\Dot{\Vec{r}}_\mathrm{sun})\cdot\Vec{S}}{c}\right]\Vec{S} \\
&-\left[\frac{GM_\mathrm{sun}}{{|\Vec{r}-\Vec{r}_\mathrm{sun}|}^3}(\Vec{r}-\Vec{r}_\mathrm{sun})+\frac{GM_\mathrm{sun}}{|\Vec{r}_\mathrm{sun}|^3}\Vec{r}_\mathrm{sun}\right] \\
&-\left(\frac{S_\mathrm{flux}A_\mathrm{g}}{m_\mathrm{g}c}\right)Q_\mathrm{pr}\frac{\Dot{\Vec{r}}-\Dot{\Vec{r}}_\mathrm{sun}}{c}-GM_\mathrm{J}R_\mathrm{J}^{2}J_{2}\triangledown\left[\frac{P_{2}(\cos{\theta})}{r^3}\right] \\
&-\sum\left[\frac{GM_\mathrm{Galilean}}{{|\Vec{r}-\Vec{r}_\mathrm{Galilean}|}^3}(\Vec{r}-\Vec{r}_\mathrm{Galilean})+\frac{GM_\mathrm{Galilean}}{|\Vec{r}_\mathrm{Galilean}|^3}\Vec{r}_\mathrm{Galilean}\right],
    \end{split}
    \label{eq:5}
\end{equation}
where $\Vec{r}$ is the radius vector of the grain, $S_\mathrm{flux}$ the solar flux at Jupiter’s distance, $A_\mathrm{g}$ the cross-section of particle, $m_\mathrm{g}$ the dust's mass, $Q_\mathrm{pr}$ the solar radiation pressure coefficient calculated by Mie calculation \citep{burns1979radiation}, $\Vec{r}_\mathrm{sun}$ the radius vector of the Sun, $c$ the speed of light, and $\Vec{S}=\frac{\Vec{r}-\Vec{r}_\mathrm{sun}}{|\Vec{r}-\Vec{r}_\mathrm{sun}|}$ the unit vector from the Sun to the dust. The variable $M_\mathrm{sun}$ is the mass of the Sun, $J_2$ the Jupiter’s zonal harmonic of degree 2, $P_2$ the second-degree Legendre function, $\theta$ the colatitude in the Jupiter-centered body-fixed frame, and $M_\mathrm{Galilean}$ the mass of Galilean satellite.

The $J_4$ term of Jupiter's gravity is omitted in our simulation, because $J_4$ ($\sim-5.87\mathrm{E}$-$4$) is about two orders of magnitude lower than $J_2$ ($\sim1.47\mathrm{E}$-$2$) \citep{durante2020jupiter} and the Jovian irregular satellites are far away from Jupiter (> 100 $R_\mathrm{J}$ as shown in Fig.~\ref{fig1}).
%since the dust from irregular satellites are far from the central planet and $J_4$ is three orders of magnitude lower than $J_2$.}
Besides, only the gravitational perturbations from the outer Galilean satellites Callisto and Ganymede are included during numerical integration in order to save the computing resources.
This simplification is reasonable since most of the dust particles are eliminated by these two moons once dust get close to their territories \cite[]{bottke2013black,haghighipour2008region}.
% bottke的文章有与Io和Europa相撞的!有的，所以为什么要删掉Io和Europa的引力!算不动

\subsection{Simulation setting}
\label{sec:Simulation program}
The escape velocities of Jupiter's irregular satellites are much smaller than their orbital velocities because of their small sizes (e.g.~escape velocity $\sim0.06\,\mathrm{km\,s^{-1}}$ vs.~orbital velocity $\sim3\,\mathrm{km\,s^{-1}}$ for Jupiter's largest irregular satellite Himalia).
Besides, the ejection velocity of impact-generated ejecta approximately follows a power law distribution \citep{housen2011ejecta}, which means that the proportion of ejecta with speeds much higher than the escape velocity of the parent moon is very small. Thus, it is expected that the initial orbital elements of most ejected particles are approximately equal to those of the parent satellites at one individual osculating orbital position of the satellites.

It should also be noted that the semimajor axes $a$, eccentricities $e$ and inclinations $i$ of Jupiter's irregular satellites evolve and vary greatly over one thousand years \cite[]{brozovic2017orbits}. In order to include this effect, 200 particles for each grain size are simulated, with different initial orbital elements, where the initial $a$, $e$, and $i$ are randomly distributed within their respective variation ranges of the corresponding parent satellite families (see Table \ref{tab:2}). Ten different values of both initial argument of pericenter $\omega$ and initial longitude of ascending node $\Omega$ are taken uniformly in the range of $[0^{\circ}, 360^{\circ}]$, because both $\omega$ and $\Omega$ are precessing with time. The initial true anomalies $f$ of ejecta are also randomly chosen within $[0^{\circ}, 360^{\circ}]$. Based on our selection of initial orbital elements, the distributions of both initial orbital velocities and initial spatial positions of dust around Jupiter cover broad ranges.

Trajectories of dust particles with broad ranges of initial orbital velocities and initial spatial positions are simulated by using a well-tested code \cite[see][]{liu2016dynamics,liu2018dust,liu2021configuration} to integrate the equation of motion shown in Section \ref{sec:The equation of dust's motion}.
%Based on the dynamical equation shown in Section \ref{sec:The equation of dust's motion}, trajectories of dust particles in the Jovian equatorial inertial frame JIF are simulated by a well-tested code \cite[see][]{liu2016dynamics,liu2018dust,liu2021configuration}.
%\textcolor{violet}{Trajectories of dust particles with broad ranges of intial orbital velocities and initial spatial positions in the Jovian equatorial inertial frame JIF are simulated by integrate the dynamical equation shown in Section \ref{sec:The equation of dust's motion} with a well-tested code \cite[see][]{liu2016dynamics,liu2018dust,liu2021configuration}.}
The orbits of dust grains are integrated until they hit a sink (impact Galilean moons, Jupiter or escape from the Jovian system).

\begin{table}
	\centering
	%\caption{This is an example table. Captions appear above each table.
	%Remember to define the quantities, symbols and units used.}
        \caption{The variation ranges of orbital elements of four satellite families over 1000 years \citep{brozovic2017orbits}.}
 
        \label{tab:2}
	\begin{tabular}{lcccr} % four columns, alignment for each
		\hline
		& Himalia & Ananke & Carme & Pasiphae\\
		\hline
		$a \left[R_\mathrm{J}\right]$ & 155-175 & 270-310 & 305-345 & 310-350\\
		$e$ & 0.1-0.3 & 0-0.5 & 0.1-0.45 & 0.1-0.7\\
		$i \left[\mathrm{deg}\right]$ & 25-30 & 140-155 & 160-170 & 140-160\\
		\hline
	\end{tabular}
\end{table}

% Note that the PR drag is multiplied by a factor related to the grain size for large particles ($r_\mathrm{g}>10{\mu}$m) in the integration in order to speed up the simulation, and the rationality of this method has been verified by \cite{hamilton2015small}.
Note that for large particles ($r_\mathrm{g}>10\,{\mu}$m), the PR drag is multiplied by a factor related to the grain size in the numerical integration in order to speed up the simulation, which reads
\begin{equation}
    \ddot{\vec{r}}_\mathrm{PR}=-C_\mathrm{PR}\left(\frac{S_\mathrm{flux}A_\mathrm{g}}{m_\mathrm{g}c}\right)Q_\mathrm{pr}\frac{\Dot{\Vec{r}}-\Dot{\Vec{r}}_\mathrm{sun}}{c}.
\end{equation}
Here $\ddot{\vec{r}}_\mathrm{PR}$ is the acceleration due to the PR drag, and $C_\mathrm{PR}$ is the grain-size-dependent enhancement factor. The similar method was used in the studies of Jupiter's gossamer rings \cite[]{burns1999formation} and Saturn's Phoebe ring \cite[]{hamilton2015small}.

\section{simulation results}
\subsection{Average dynamical lifetimes of particles}
\label{sec:lifetime}
\begin{figure}
	\includegraphics[width=\columnwidth]{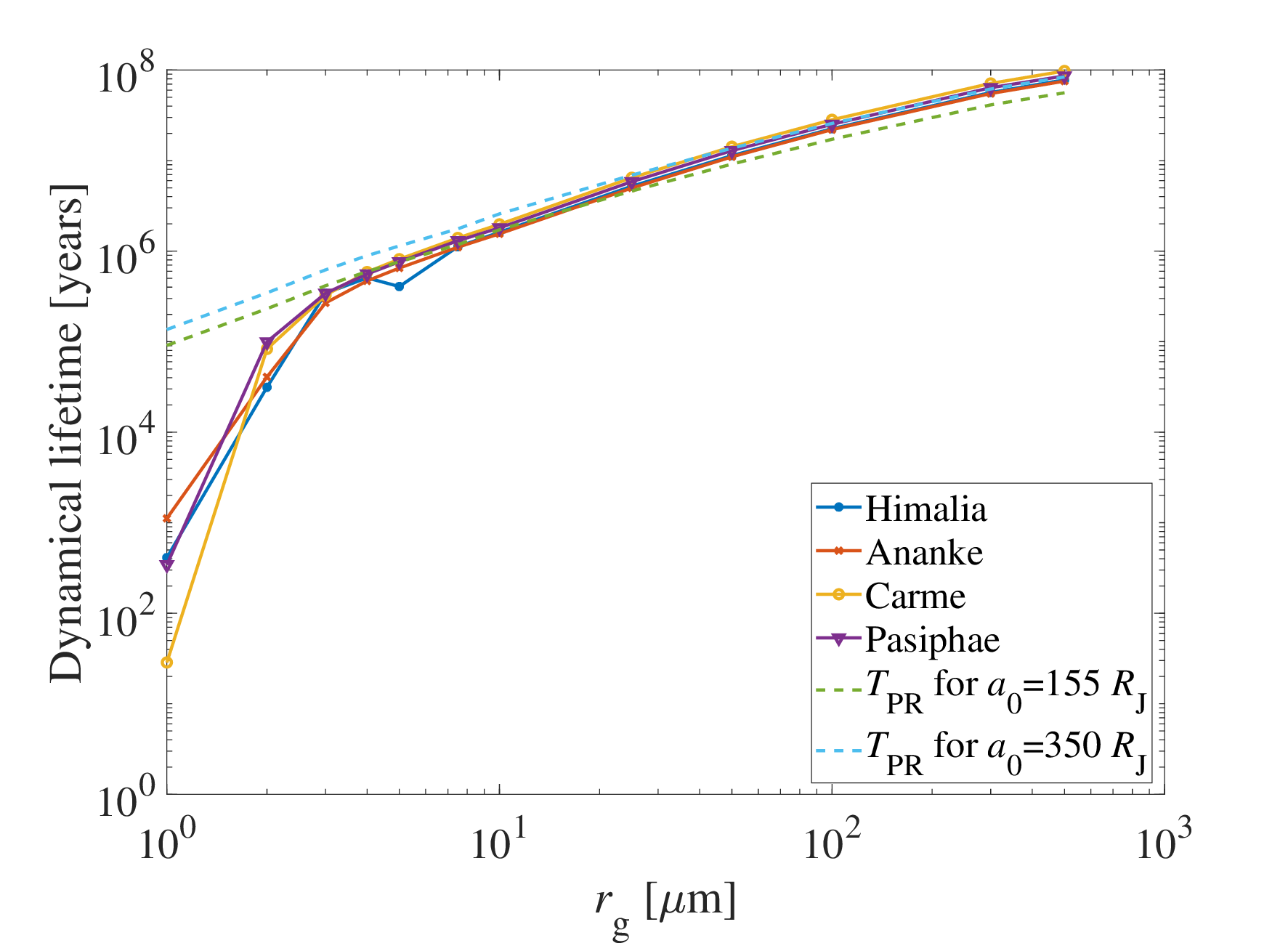}
    \caption{Average dynamical lifetimes obtained from numerical simulations for dust with different sizes originating from different families of satellites (solid lines with symbols). The dashed lines denote the inward migration time $T_\mathrm{PR}$ estimated analytically by Equation (\ref{eq:12}) for dust with the smallest ($a_0=155\,R_\mathrm{J}$) and the largest ($a_0=350\,R_\mathrm{J}$) initial semimajor axes used in our numerical simulations (see Table \ref{tab:2}).}
    \label{fig2}
\end{figure}

The average dynamical lifetimes of dust with different sizes (Fig.~\ref{fig2}) are easily obtained from our numerical simulations.
The dynamical lifetimes of 1 $\mu$m grains are short, which are on the order of 100 to 1000 years. As the grain size increases, its lifetime gets longer.
% The growth rate of lifetime between 1 and 3 $\mu$m is fast but instead is flat for dust larger than 3 $\mu$m.
The growth rate of dynamical lifetimes between 1 and 3 $\mu$m is faster, but slower for dust larger than 3 $\mu$m.
% After size of grain up to 10 $\mu$m, particles can survive in space over a million years.
Large particles ($r_\mathrm{g}>10\,{\mu}$m) can survive in space over a million years.
% Note that the decline of lifetime between 4 and 5 $\mu$m for dust from the Himalia family is unexpected, which we speculate it is due to the coupling of radiation pressure and solar gravity according to some numerical tests.
% which we speculate it is due to the coupling of radiation pressure and solar gravity according to some numerical tests.
%Note that the decrease of lifetime between 4 and 5 $\mu$m for dust from the Himalia family is unexpected, which will be investigated in the future study.

\begin{figure}
	\includegraphics[width=\columnwidth]{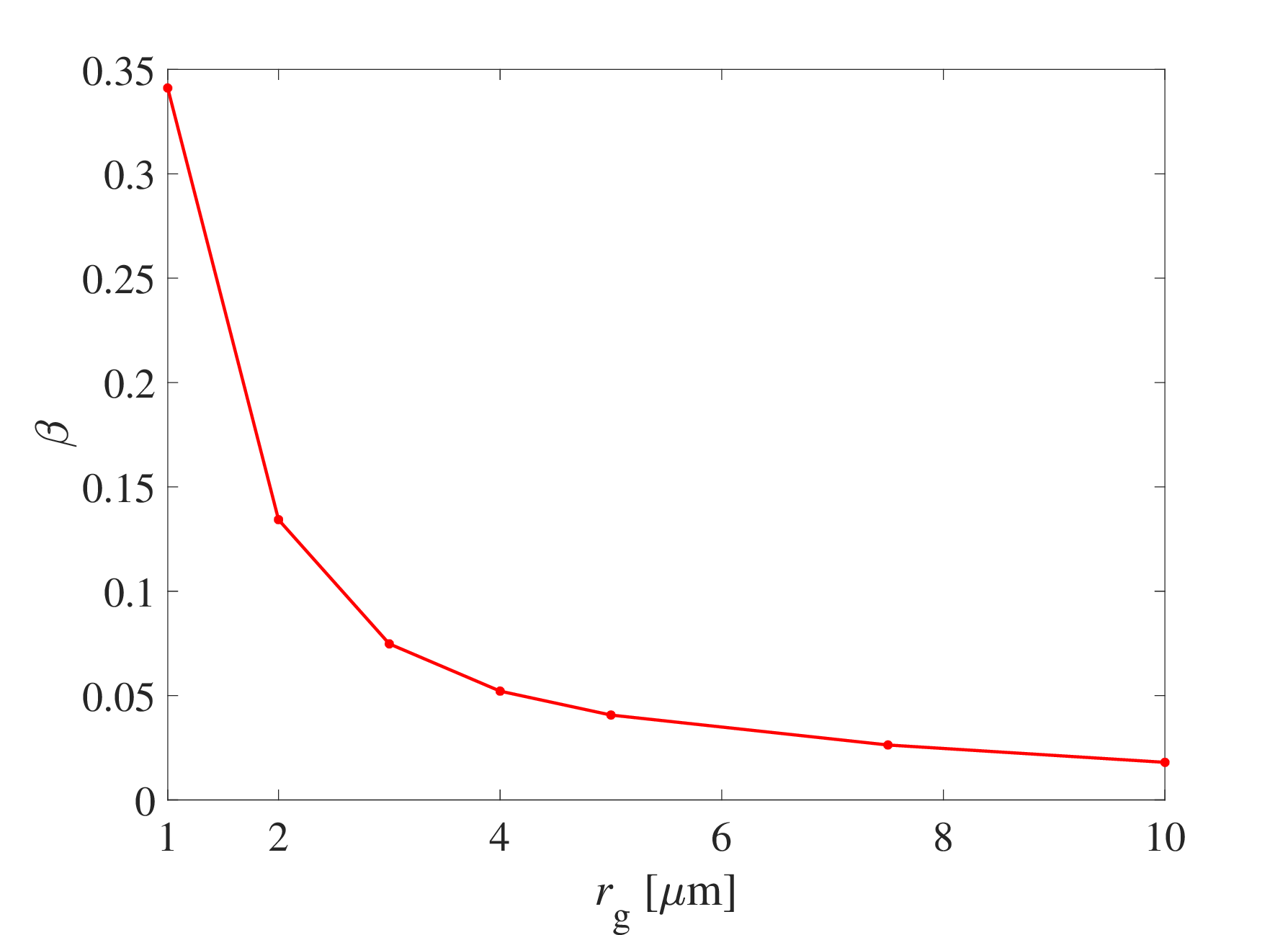}
    \caption{The ratio of solar radiation pressure relative to solar gravity as a function of grain size.}
    \label{fig3}
\end{figure}

% Ignoring the curious behavior of 5 $\mu$m particles from \textcolor{red}{the} Himalia family, the ratio of radiation pressure to the Sun’s gravity \cite[]{burns1979radiation},
The effect of solar radiation pressure is strong for dust smaller than a few micrometers.
The ratio of solar radiation pressure to solar gravity reads \cite[]{burns1979radiation}
\begin{equation}
     \beta=\frac{3Q_\mathrm{s}Q_\mathrm{pr}\mathrm{AU}^2}{4GM_\mathrm{sun}c}\left(\frac{1\,\mathrm{g\,cm^{-3}}}{{\rho}}\right)\left(\frac{1\,\mathrm{cm}}{r_\mathrm{g}}\right),
        \label{eq:6}
\end{equation}
where $Q_\mathrm{s}$ is the solar flux at one $\mathrm{AU}$ (astronomical unit).
As shown in Fig.~\ref{fig3}, the value of $\beta$ is largest at 1 $\mu$m and decreases as the grain size increases, where the rate of decay is faster between 1 and 3 $\mu$m and slower for dust larger than 3 $\mu$m.
% In physics, a larger value of $\beta$ implies the stronger influence of radiation pressure, which causes a larger amplitude of eccentricity oscillation.
The larger value of $\beta$ implies the stronger influence of solar radiation pressure, which causes a larger amplitude of the eccentricity oscillation \cite[]{hamilton1996circumplanetary}.
% Therefore, \textcolor{red}{the} high value of $\beta$ at 1 $\mu$m indicates that these particles can easily strike into Jupiter or escape from Jovian system due to their extremely large eccentricities, which corresponds to their short life.
Therefore, 1 $\mu$m particles are quickly removed by striking into Jupiter or escaping from the Jovian system due to their extremely large eccentricities, and thus have a short dynamical lifetime.
% As the effect of the radiation pressure diminishes, the dust's lifetime increases.
As the solar radiation pressure diminishes, the dust's lifetime increases.

\begin{figure}
	\includegraphics[width=\columnwidth]{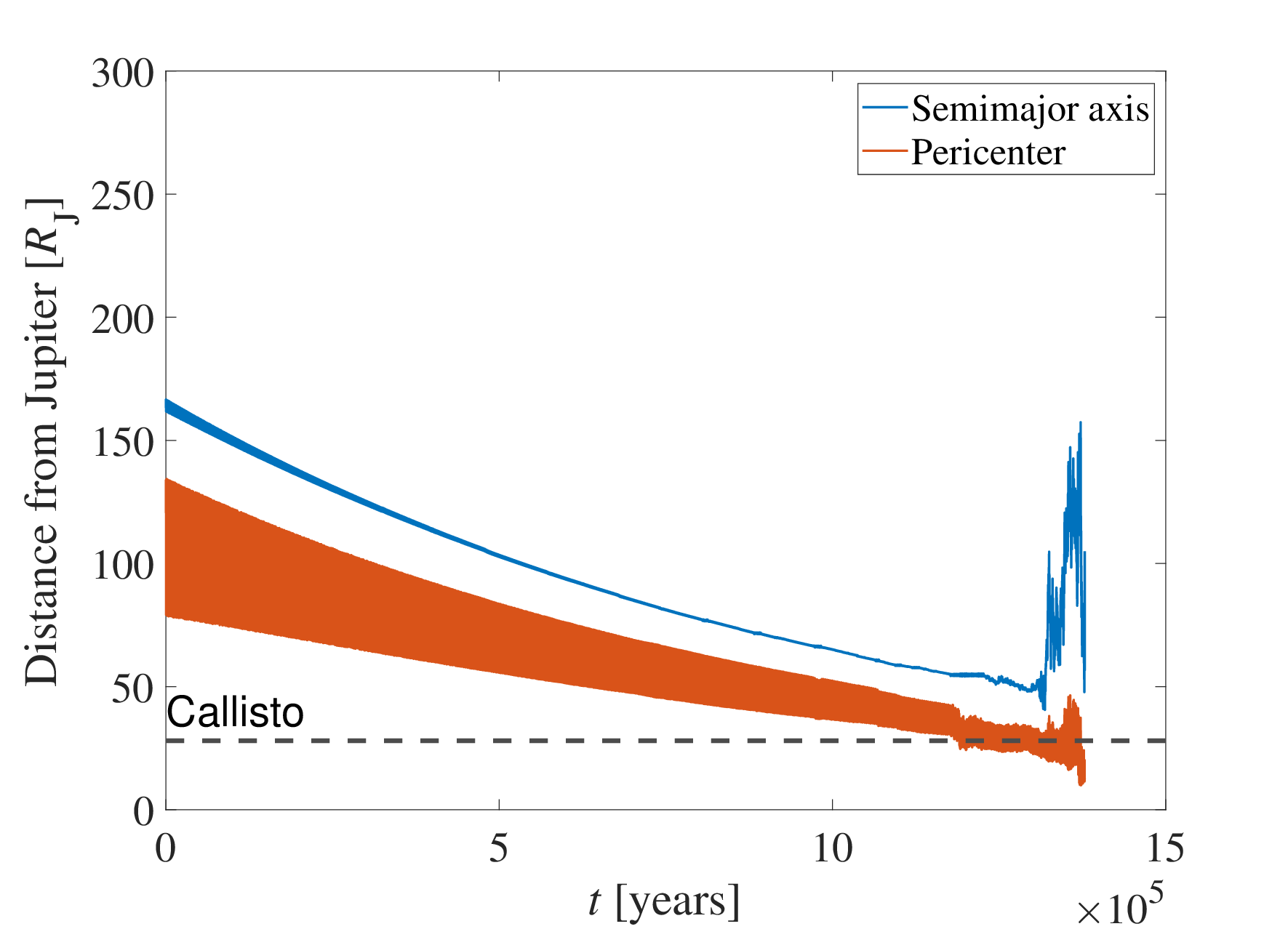}
    \caption{Typical evolution of semimajor axis and pericenter of a 10 $\mu$m particle from the Himalia family. The black dashed line denotes the orbital distance of Callisto.}
    \label{fig4}
\end{figure}

% Actually, the lifetime of dust larger than a few micrometers is approximately equal to the time it takes for the dust to transport inward to the central planet due to the Poynting-Robertson drag, which can be estimated by \cite[]{burns1979radiation}
The dynamical lifetime of dust larger than a few micrometers is approximately equal to the inward migration time of dust from irregular moons to the regions of Galilean satellites due to the PR drag, which can be estimated by \cite[]{burns1979radiation, krivov2002dust}
\begin{equation}
    T_\mathrm{PR}=\tau_\mathrm{PR}\ln{\left(\frac{a_0}{a_1}\right)}.
        \label{eq:12}
\end{equation}
% 28\,R_\mathrm{J}
Here $a_0$ is the initial semimajor axis of dust particles around Jupiter, and $a_1$ is the outer boundary of the influence zone of Callisto, which is about $28\,R_\mathrm{J}$ according to \citet{haghighipour2008region}. The variable $\tau_\mathrm{PR}$ is the exponential decay time \cite[]{burns1979radiation},
\begin{equation}
     \tau_\mathrm{PR}=9.3{\times}\frac{10^{6}R^{2}}{Q_\mathrm{pr}}\left(\frac{{\rho}}{1\,\mathrm{g\,cm^{-3}}}\right)\left(\frac{r_\mathrm{g}}{1\,\mathrm{cm}}\right),
        \label{eq:13}
\end{equation}
% where $R$ is the Jupiter’s heliocentric distance in $\mathrm{AU}$ and other parameters have been mentioned before.
where $R$ is the heliocentric distance of dust particles in AU. The analytical estimations of $T_\mathrm{PR}$ for dust with the smallest ($a_0=155\,R_\mathrm{J}$) and the largest ($a_0=350\,R_\mathrm{J}$) initial semimajor axes used in our numerical simulations (see Table \ref{tab:2}) are overplotted in Fig.~\ref{fig2} for comparison. Obviously, the analytical results estimated by Equation (\ref{eq:12}) for particles larger than $3\,\mu$m are basically consistent with the average dynamical lifetimes calculated from numerical simulations; while for particles smaller than $3\,{\mu}$m, Equation (\ref{eq:12}) cannot give satisfactory estimations of lifetimes because these small particles are subjected to stronger solar radiation pressure.
%where $R$ is the Jupiter’s heliocentric distance in $\mathrm{AU}$. According to this formula, the inward migration time for a 10 $\mu$m particle is about one million years, which is basically consistent with the average lifetime calculated from our simulation.
Fig.~\ref{fig4} shows a typical evolution of semimajor axis and pericenter of a 10 $\mu$m particle.
The semimajor axis decays slowly due to the PR drag.
After dust’s pericenter reachs the regions of Galilean satellites, the orbit of the grain becomes unstable under the gravitational perturbations of Galilean satellites and quickly hits a sink.

%The average life span of particles larger than 10 $\mu$m are not shown in Fig.~\ref{fig2},
%because the simulation of these dust particles are accelerated,
%and therefore, the lifetime cannot be obtained directly.
%Nevertheless, after weighting the enhancement factor of PR drag with the simulated evolution time of dust, the lifetime is well matched with the estimation by equation (7).

\subsection{Dust's final fate}
\label{sec:final fate}
% Although the average lifetimes of dust particles ejected from four irregular satellite groups is similar,
Although the average dynamical lifetimes of dust particles ejected from four irregular satellite groups are close to each other,
% the final fate of grains are fairly different, which are summaries in Tables $3-6$.
the final fates of grains, which are summarized in Tables \ref{tab:3}-\ref{tab:6}, are fairly different.

\begin{table}
	\centering
	%\caption{This is an example table. Captions appear above each table.
	%Remember to define the quantities, symbols and units used.}
        \caption{The proportion of different destinations of dust particles from the prograde Himalia family.}
        \resizebox{\columnwidth}{!}{
        \label{tab:3}
	\begin{tabular}{lccccccr} % four columns, alignment for each
		\hline
		$r_\mathrm{g}(\mathrm{\mu m})$ & Escape & Hit Jupiter & Hit Io & Hit Europa & Hit Ganymede & Hit Callisto\\
		\hline
		1 & 0.125 & 0.87 & 0.005 & 0 & 0 & 0\\
		2 & 0.355 & 0.365 & 0.025 & 0.145 & 0.025 & 0.085\\
		3 & 0.26 & 0.015 & 0.015 & 0.125 & 0.14 & 0.445\\
            4 & 0.24 & 0.06 & 0.01 & 0.075 & 0.12 & 0.495\\
            5 & 0.385 & 0.145 & 0.005 & 0.045 & 0.07 & 0.35\\
            7.5 & 0.115 & 0.095 & 0.01 & 0.14 & 0.2 & 0.44\\
            10 & 0.125 & 0.045 & 0.005 & 0.095 & 0.235 & 0.495\\
            25 & 0.155 & 0.045 & 0.005 & 0.105 & 0.13 & 0.56\\
            50 & 0.205 & 0.025 & 0 & 0.095 & 0.2 & 0.475\\
            100 & 0.165 & 0.03 & 0.01 & 0.135 & 0.175 & 0.485\\
            300 & 0.17 & 0.025 & 0.01 & 0.11 & 0.235 & 0.45\\
            500 & 0.12 & 0.045 & 0 & 0.055 & 0.18 & 0.6\\
		\hline
	\end{tabular}}
\end{table}

\begin{table}
	\centering
	%\caption{This is an example table. Captions appear above each table.
	%Remember to define the quantities, symbols and units used.}
        \caption{The proportion of different destinations of dust particles from the retrograde Ananke family.}
        \resizebox{\columnwidth}{!}{
        \label{tab:4}
	\begin{tabular}{lccccccr} % four columns, alignment for each
		\hline
		$r_\mathrm{g}(\mathrm{\mu m})$ & Escape & Hit Jupiter & Hit Io & Hit Europa & Hit Ganymede & Hit Callisto\\
		\hline
		1 & 0.9 & 0.085 & 0 & 0.005 & 0.01 & 0\\
		2 & 0.31 & 0.445 & 0 & 0 & 0.115 & 0.13\\
		3 & 0.05 & 0.095 & 0 & 0 & 0.22 & 0.635\\
            4 & 0 & 0.02 & 0.005 & 0 & 0.115 & 0.86\\
            5 & 0.005 & 0.005 & 0 & 0 & 0.065 & 0.925\\
            7.5 & 0 & 0 & 0 & 0 & 0.035 & 0.965\\
            10 & 0 & 0 & 0 & 0 & 0.045 & 0.955\\
            25 & 0 & 0 & 0 & 0 & 0.01 & 0.99\\
            50 & 0 & 0 & 0 & 0 & 0.01 & 0.99\\
            100 & 0 & 0 & 0 & 0 & 0.005 & 0.995\\
            300 & 0 & 0 & 0 & 0 & 0.015 & 0.985\\
            500 & 0 & 0 & 0 & 0 & 0.005 & 0.995\\
		\hline
	\end{tabular}}
\end{table}

\begin{table}
	\centering
	%\caption{This is an example table. Captions appear above each table.
	%Remember to define the quantities, symbols and units used.}
        \caption{The proportion of different destinations of dust particles from the retrograde Carme family.}
        \resizebox{\columnwidth}{!}{
        \label{tab:5}
	\begin{tabular}{lccccccr} % four columns, alignment for each
		\hline
		$r_\mathrm{g}(\mathrm{\mu m})$ & Escape & Hit Jupiter & Hit Io & Hit Europa & Hit Ganymede & Hit Callisto\\
		\hline
		1 & 0.805 & 0.195 & 0 & 0 & 0 & 0\\
		2 & 0.34 & 0.175 & 0 & 0.01 & 0.2 & 0.275\\
		3 & 0.2 & 0 & 0 & 0 & 0.045 & 0.755\\
            4 & 0.08 & 0.005 & 0 & 0 & 0.015 & 0.9\\
            5 & 0.03 & 0 & 0 & 0 & 0.01 & 0.96\\
            7.5 & 0 & 0.005 & 0 & 0 & 0 & 0.995\\
            10 & 0.01 & 0.01 & 0 & 0 & 0 & 0.98\\
            25 & 0.01 & 0.005 & 0 & 0 & 0.005 & 0.98\\
            50 & 0.015 & 0 & 0 & 0 & 0 & 0.985\\
            100 & 0 & 0 & 0 & 0 & 0.01 & 0.99\\
            300 & 0.005 & 0 & 0 & 0 & 0.005 & 0.99\\
            500 & 0.005 & 0 & 0 & 0 & 0.005 & 0.99\\
		\hline
	\end{tabular}}
\end{table}

\begin{table}
	\centering
	%\caption{This is an example table. Captions appear above each table.
	%Remember to define the quantities, symbols and units used.}
        \caption{The proportion of different destinations of dust particles from the retrograde Pasiphae family.}
        \resizebox{\columnwidth}{!}{
        \label{tab:6}
	\begin{tabular}{lccccccr} % four columns, alignment for each
		\hline
		$r_\mathrm{g}(\mathrm{\mu m})$ & Escape & Hit Jupiter & Hit Io & Hit Europa & Hit Ganymede & Hit Callisto\\
		\hline
		1 & 0.785 & 0.215 & 0 & 0 & 0 & 0\\
		2 & 0.45 & 0.105 & 0 & 0. & 0.16 & 0.285\\
		3 & 0.22 & 0.045 & 0 & 0 & 0.125 & 0.61\\
            4 & 0.155 & 0.035 & 0 & 0 & 0.065 & 0.745\\
            5 & 0.11 & 0.015 & 0 & 0 & 0.025 & 0.85\\
            7.5 & 0.08 & 0.01 & 0 & 0 & 0.04 & 0.87\\
            10 & 0.08 & 0.015 & 0 & 0 & 0.03 & 0.875\\
            25 & 0.08 & 0.01 & 0 & 0 & 0.005 & 0.905\\
            50 & 0.085 & 0 & 0 & 0 & 0.005 & 0.91\\
            100 & 0.09 & 0.015 & 0 & 0 & 0.015 & 0.88\\
            300 & 0.07 & 0.005 & 0 & 0 & 0.01 & 0.915\\
            500 & 0.09 & 0.02 & 0 & 0 & 0.01 & 0.88\\
		\hline
	\end{tabular}}
\end{table}

% As analyzed in Section 4.1, 1 $\mu$m particles would quickly hit the planet or escape from the Jovian system due to the strong effect of solar radiation pressure,
As analyzed in Section \ref{sec:lifetime}, 1 $\mu$m particles are strongly affected by solar radiation pressure and quickly hit the planet or escape from the Jovian system.
% no matter which moon families that dust originated from.
% When particles are small ($r_\mathrm{g}<5{\mu}$m), the fractions of escape and hitting the planet decrease as the grain size increases,
Generally, when particles are small ($r_\mathrm{g}<5\,{\mu}$m), the fractions of dust particles that impact Jupiter and flee the Jovian system decrease as the grain size increases, because the effect of solar radiation pressure decreases.
% as a consequence of the solar radiation is becoming less important.
%as a consequence of the reduced importance of radiation pressure.
% For particles from the Carme and Ananke groups, these fractions drop to almost zero after the dust reaches 5 $\mu$m;
For particles from the Carme and Ananke groups, these fractions drop to almost zero as the grain size increases to 5 $\mu$m;
% while for large dust ($r_\mathrm{g}>5{\mu}$m) from the Pasiphae and Himalia families, they add up to approximately 10$\%$ and 20$\%$ respectively, regardless of the grain size.
while for large particles ($r_\mathrm{g}>5\,{\mu}$m) from the Pasiphae and the Himalia families, the sums of the two fractions are approximately 10$\%$ and 20$\%$, respectively, regardless of the grain size.

\begin{figure}
	\includegraphics[width=\columnwidth]{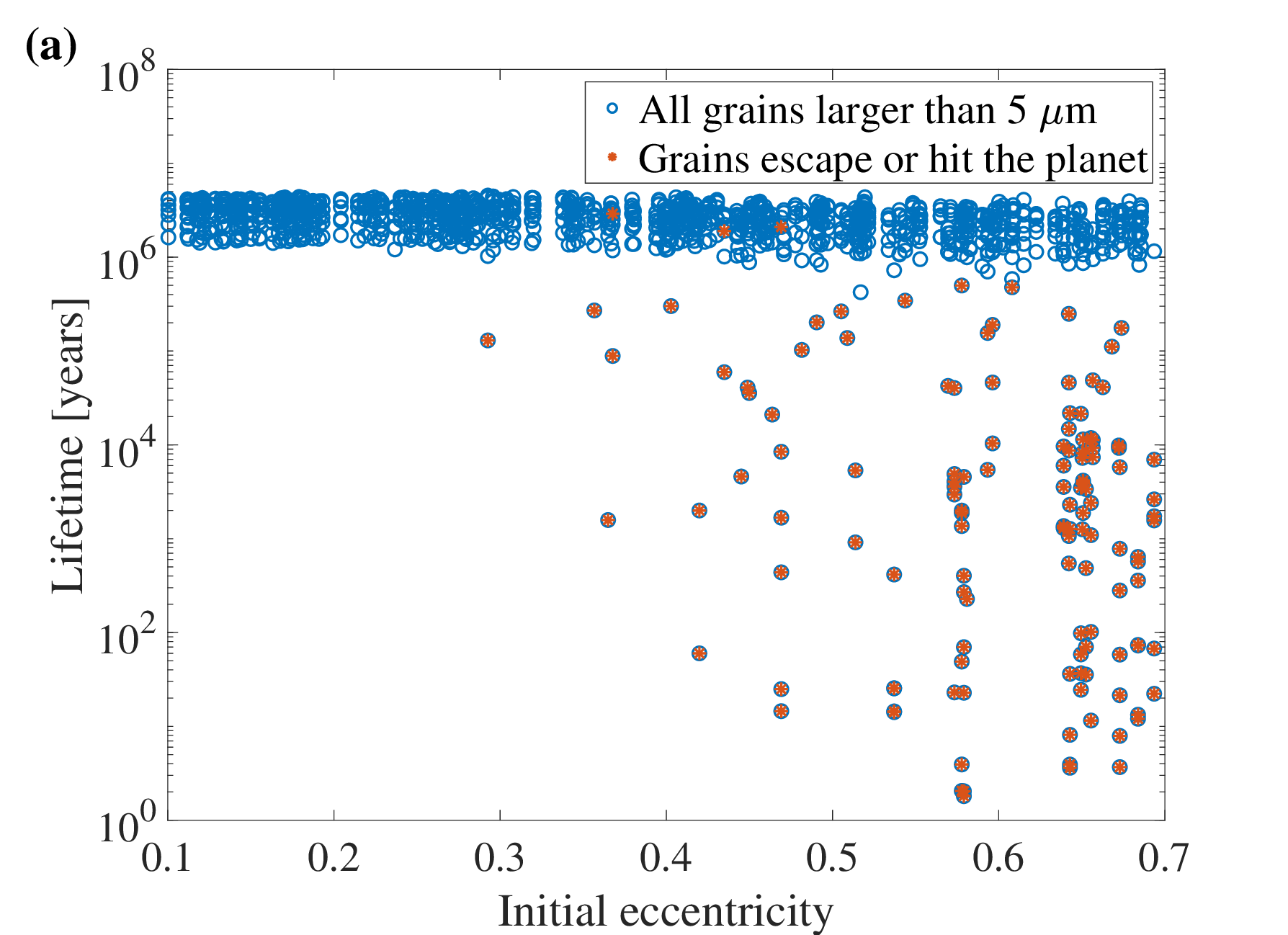}
         \includegraphics[width=\columnwidth]{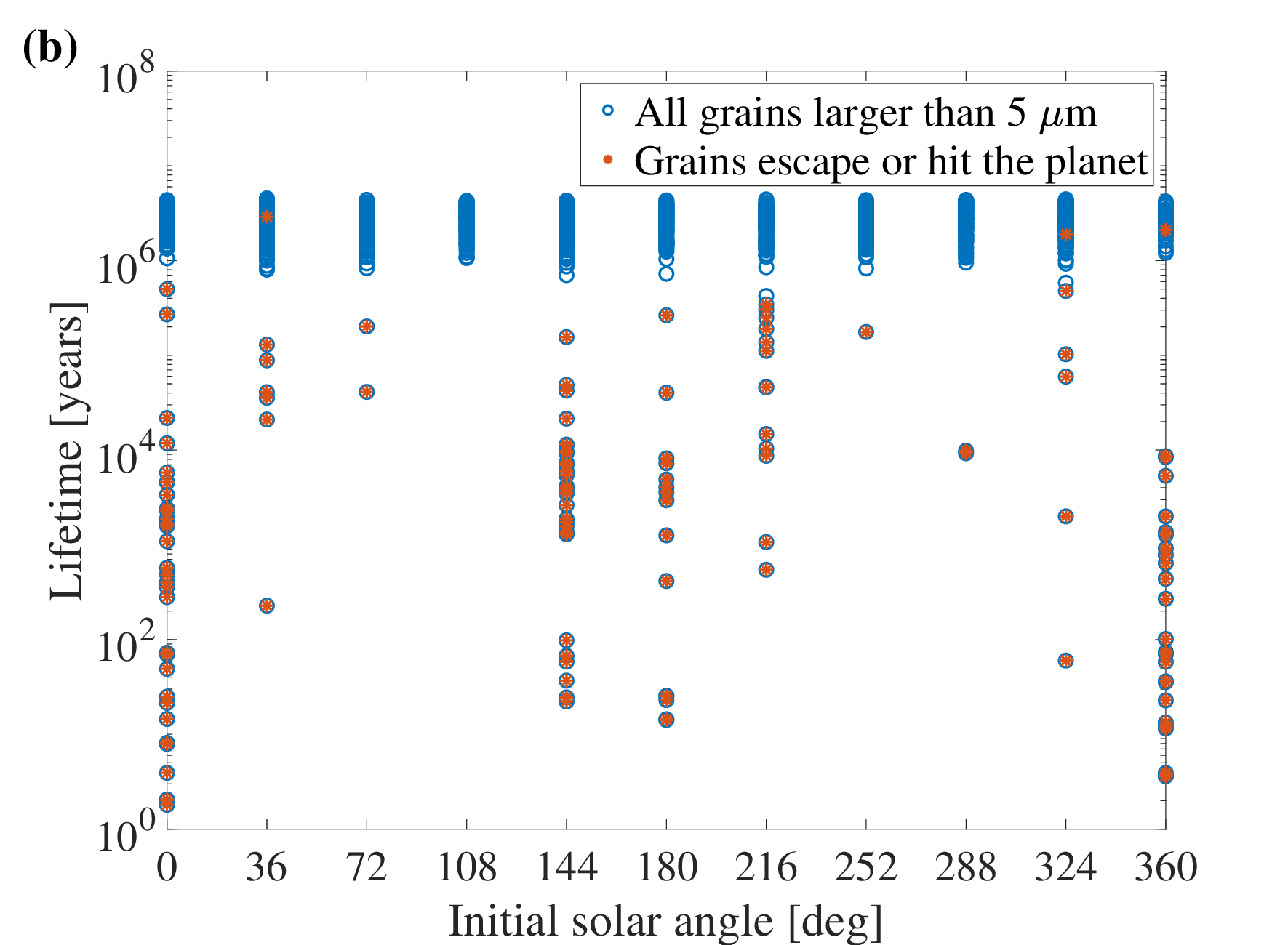}
    %\caption{(a) The initial eccentricity of dust larger than 5 $\mu$m from the Pasiphae group versus their lifetimes. The blue dots contain all grains, and the asterisk is used to mark grains which escape or hit the planet. (b) The initial solar angle versus lifetime for the same dust in (a). The usage of dots and asterisk are the same as (a).}
    \caption{(a) The initial eccentricity of dust larger than 5 $\mu$m from the Pasiphae group versus their dynamical lifetime. The blue dots denote all grains, and the red asterisks represent the grains that escape from the Jovian system or hit the planet. (b) Same as (a), but for the initial solar angle versus dynamical lifetime.}
    \label{fig5}
\end{figure}

\begin{figure}
	\includegraphics[width=\columnwidth]{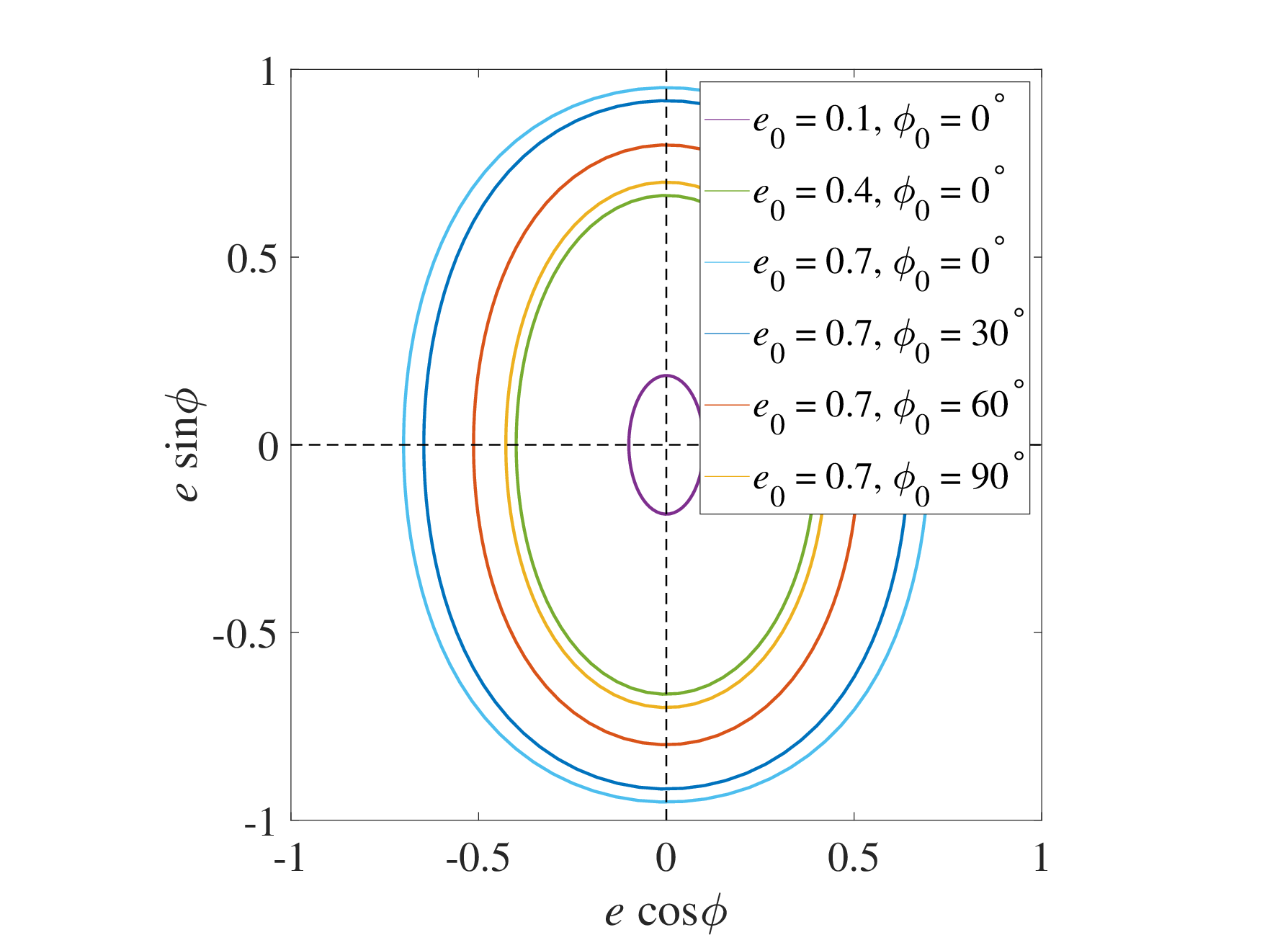}
    \caption{Phase portrait for a 100 $\mu$m particle from the Pasiphae group. Different colored curves correspond to the evolution of dust with different initial eccentricities and solar angles.}
    \label{fig6}
\end{figure}

% Fig.~\ref{fig5} shows the initial eccentricity and solar angle of all the large particles ($r_\mathrm{g}>5{\mu}$m) from Pasiphae versu their lifetime.
For large particles ($r_\mathrm{g}>5\,{\mu}$m) from the Pasiphae family, the relations between the dynamical lifetimes of dust and their initial eccentricities $e$ and solar angles $\phi$, are shown in Fig.~\ref{fig5}, where $\phi$ is approximately the angle between the radius vectors from Jupiter to the Sun and to the pericenter of dust \citep{hamilton1993motion}.
% Most of the particles escape or impacting Jupiter are short-lived with a large initial eccentricity and a solar angle close to $0^{\circ}$ or $180^{\circ}$.
Most of the particles that escape from the Jovian system or impact Jupiter are short-lived, with large initial eccentricities and initial solar angles close to $0^{\circ}$ or $180^{\circ}$.
This can be understood through the conserved integral of the dust’s motion derived by \cite{hamilton1996circumplanetary}.
% For large particles from the Pasiphae group, solar gravity is the dominant perturbation, and thus the integral of motion, which is named the “Hamiltonian”, is written as
For large particles from the Pasiphae group, solar gravity is the dominant perturbation, and thus the integral of motion, known as the “Hamiltonian”, is written as
\begin{equation}
     H=-\sqrt{1-e^2}+\frac{1}{2}C_\mathrm{sg}e^2\left(1+5\cos{2{\phi}}\right),
        \label{eq:8}
\end{equation}
where $C_\mathrm{sg}$ is the parameter describes the strength of solar gravity.
Based on the conserved “Hamiltonian”, the subsequent evolution of the solar angle and eccentricity for any given initial value can be followed analytically, as shown in Fig.~\ref{fig6}.
% Compared to the other phase trajectories, the orbit of dust with a large initial eccentricity and a initial solar angle close to $0^{\circ}$ or $180^{\circ}$ becomes most eccentric under the disturbing by solar gravity, which leads to escape or collide with Jupiter.
% Compared to the other phase trajectories, the orbits of dust with large initial eccentricities and initial solar angles close to $0^{\circ}$ or $180^{\circ}$ becomes most eccentric under the perturbation of solar gravity, which leads to escape from the Jovian system or collision with Jupiter.
The orbit of dust with large initial eccentricity $e_0$ and initial solar angle $\phi_0$ near $0^{\circ}$ or $180^{\circ}$ gets the largest eccentricity close to 1 under the perturbation of solar gravity, which results in the escape from the Jupiter system or collision with the Jupiter.

\begin{figure}
	\includegraphics[width=\columnwidth]{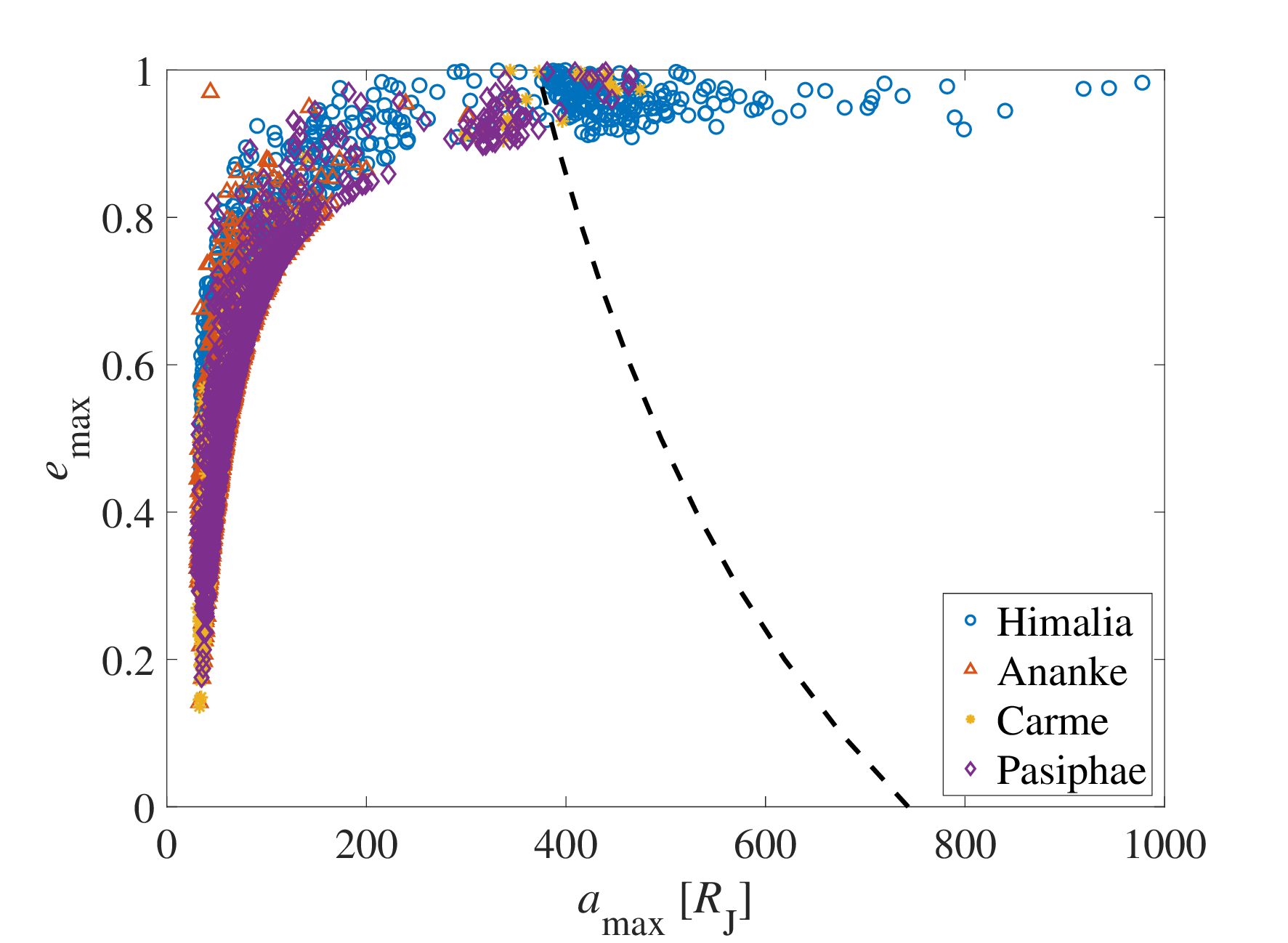}
    %\caption{The maximum semimajor axes and eccentricity of dust \textcolor{red}{particles} larger than 10 $\mu$m after their pericenter has got into the influence zones of Galilean satellites. The black dashed line represents the condition for dust to escape, that is, the dust on the left of the line cannot escape, while the dust located at the right of the line satisfies the condition for leaving the Jovian system.}
    \caption{The maximum semimajor axes and eccentricities of large particles ($r_\mathrm{g}>5\,{\mu}$m) after their pericenters arrive at the regions of Galilean satellites. The black dashed line represents the boundary of escape. Dust particles located on the left of the line cannot escape from the Jovian system, while the dust particles located at the right of the line satisfy the condition for leaving the Jovian system.}
    \label{fig7}
\end{figure}

% Unlike the dust from \textcolor{red}{the} Pasiphae family, the reason for large particles from \textcolor{red}{the} Himalia group that hitting the planet or being expelled is the gravitational perturbations induced by Galilean satellites rather than their initial orbital elements.
Unlike the dust from the retrograde Pasiphae family, the reason for large particles from the prograde Himalia group to hit the planet or be expelled is the gravitational perturbations induced by Galilean satellites rather than their initial orbital elements.
%Fig.~\ref{fig7} shows the maximum semimajor axes and eccentricities of all the large particles ($r_\mathrm{g}>5{\mu}$m) from four satellite families after their pericenter have got into the influence zones of Galilean satellites, which allow us to conclude that the prograde particle is much more easily excited into extremely eccentric and remote orbit by encounter with the regular moons, which result for their escape or hitting the planet.
Fig.~\ref{fig7} shows the maximum semimajor axes and eccentricities of all the large particles ($r_\mathrm{g}>5\,{\mu}$m) from four satellite families after their pericenters reach the regions of Galilean satellites.
% Compared to the retrograde particles, prograde dust grains are much more easily excited into extremely eccentric and remote orbit by encounter with the Galilean moons, which \textcolor{red}{leads to} escape from the Jovian system or impact Jupiter.
Compared to the retrograde satellite ejecta, dust grains from the prograde Himalia group are much more easily excited into extremely eccentric and remote orbit by encounter with the Galilean moons, which leads to escape from the Jovian system or impact Jupiter.
% This difference of stability between retrograde and prograde particles is also reflected in the result of \cite{bottke2013black} and \cite{haghighipour2008region}, but the dynamical cause is not clear.
This difference between the stabilities of retrograde and prograde satellite ejecta was also shown in the results of \cite{bottke2013black} and \cite{haghighipour2008region}.

\begin{figure}
	\includegraphics[width=\columnwidth]{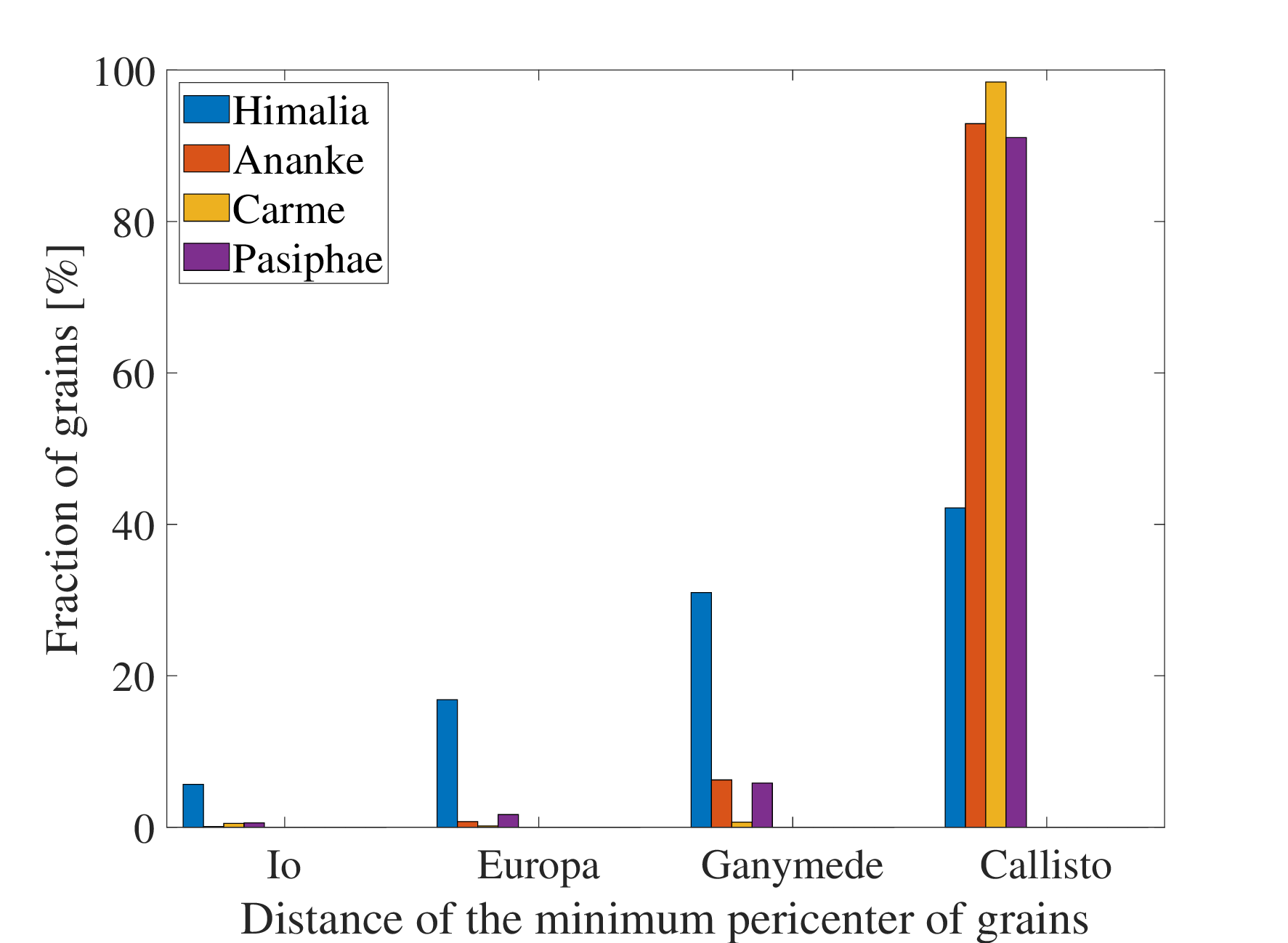}
    \caption{
    %The positions of the minimum pericenters of dust after they arrive at the territories of Galilean moons. The orbital distances of four Galilean satellites are also shown. 
    The fraction of the minimum pericenters of dust grains falling between different Galilean satellites.}
    \label{fig8}
\end{figure}

The collision probabilities with the four Galilean satellites for dust from the prograde and retrograde satellite families also vary significantly.
% Dust from retrograde satellite families is hard to hit the Io and Europa regardless of grain size, which only 3 and 1 particles respectively impact these two moons in our simulation.
Dust from the retrograde families of satellites is hard to hit Io and Europa regardless of the grain size, where the fractions are only 0.005$\%$ and 0.015$\%$ for particles hitting these two moons in our numerical simulations, respectively.
%Only 0.005$\%$ and 0.015$\%$ of \textcolor{red}{retrograde} particles impact these two moons in our simulation, respectively.
In general, most of the retrograde satellite ejecta collide with Callisto ($>90\%$ for particles larger than a few micrometers), and the survivors that bypass Callisto are mainly swept up by Ganymede.
For dust from the prograde Himalia family, the opportunities to hit the four Galilean moons are about 1$\%$, 10$\%$, 20$\%$ and 50$\%$, respectively.
% 之前3.1节末尾说只考虑Callisto和Ganymede!还是说只考虑了Io和Europa的sink，但是没考虑这个两个的gravity!考虑sink，不考虑gravity，后面补个讨论吧,在本段最后一句
%Here we recognize the different impact chances between prograde and retrograde particles by the minimum pericenters of dust after they reach the territories of Galilean mooons, because the collision is only possible when the particle’s orbital pericenter is below the moon’s semimajor axis.
Here the different collision probabilities of dust from the prograde and retrograde irregular moons with the Galilean satellites are investigated by analyzing the minimum pericenters of dust particles after they reach the territories of Galilean moons, because the collision is only possible when the particle’s orbital pericenter is smaller than the moon’s semimajor axis.
As shown in Fig.~\ref{fig8}, more than 90$\%$ of the initial retrograde dust’s minimum pericenters can only reach the distance between Callisto and Ganymede, and very a few particles can get inside the orbits of Io and Europa (0.5$\%$ and 1$\%$, respectively); while for grains from the prograde Himalia family, the fractions of the minimum pericenters located at four intervals are 5$\%$, 16$\%$, 31$\%$ and 42$\%$.
In fact, this is still the consequence of the stability distinction induced by the gravitational perturbations of Galilean satellites, which implies the different mechanism for dust striking the Galilean moons, i.e., 
dust particles from the prograde satellites hit the Galilean moons by getting into unstable orbit, while dust particles from the retrograde moons are slowly drifting by PR drag.
Based on our result, the gravitational perturbations of Io and Europa which are neglected in our simulation have an influence on the evolution of only a small fractions of dust from the prograde Himalia family.%, which may cause a slight inaccuracy in our results.

\subsection{Number density in the Jovian system}
\label{sec:Number density}
Integrating the trajectories of dust particles over their initial size distribution, the steady-state number density of dust ejected from Jupiter’s irregular moons can be estimated as follows.
The Circum-Jovian region is first divided into a cylindrical grid system with each grid cell indexed by $i_\mathrm{cell}$, $j_\mathrm{cell}$ and $k_\mathrm{cell}$.
% Then using the position of the dust’s trajectory, which is stored at equidistant time step ${\Delta}t$, to count the number of times the dust passes the grid cell ($i_\mathrm{cell},j_\mathrm{cell},k_\mathrm{cell}$), \textcolor{red}{which is} expressed as $\tilde{n}(i_\mathrm{cell},j_\mathrm{cell},k_\mathrm{cell})$.
Then the points in the dust’s trajectory, which are stored at equidistant time step ${\Delta}t$, are used to count the number of times that the dust particles with grain size $r_\mathrm{g}$ pass the grid cell ($i_\mathrm{cell},j_\mathrm{cell},k_\mathrm{cell}$), which is expressed as $\tilde{n}(i_\mathrm{cell},j_\mathrm{cell},k_\mathrm{cell}; r_\mathrm{g})$.
The readers are referred to \cite{liu2016dynamics} for more details about the gridding and storage method.
After some algebra, the estimation of number density of dust in the grid cell reads

\begin{equation}
\label{equ:numberdensity}
\resizebox{\columnwidth}{!}{$
\begin{split}
n(i_\mathrm{cell},j_\mathrm{cell},k_\mathrm{cell})=
\int_{r_\mathrm{min}}^{r_\mathrm{max}}\frac{\tilde{n}(i_\mathrm{cell},j_\mathrm{cell},k_\mathrm{cell}; r_\mathrm{g})}{V(i_\mathrm{cell},j_\mathrm{cell},k_\mathrm{cell})n_\mathrm{start}}{\Delta}t N(r_\mathrm{g})\mathrm{d}r_\mathrm{g},
\end{split}
$}
\end{equation}
%where $V(i_\mathrm{cell},j_\mathrm{cell},k_\mathrm{cell})$ is the volume of the grid cell, the summation notation is used to add up the $\tilde{n}(i_\mathrm{cell},j_\mathrm{cell},k_\mathrm{cell}; r_\mathrm{g})$ of grains with same size, and $n_\mathrm{start}$ is the number of simulated particles of each size.
where $V(i_\mathrm{cell},j_\mathrm{cell},k_\mathrm{cell})$ is the volume of the grid cell, and $n_\mathrm{start}$ is the number of simulated particles for each size. The result based on Equation (\ref{equ:numberdensity}) is presented in a rotating frame ($Ox_\mathrm{sun}y_\mathrm{sun}z_\mathrm{sun}$) that keeps the $x$-axis pointing to the Sun, in order to observe the number density distribution relative to the Sun. The $z$-axis of the $Ox_\mathrm{sun}y_\mathrm{sun}z_\mathrm{sun}$ frame is normal to the Jovian orbital plane, and the $y$-axis is completed by right-hand rule.

\begin{figure}
	\includegraphics[width=\columnwidth]{Fig10a.eps}
        \includegraphics[width=\columnwidth]{Fig10b.eps}
    \caption{(a) The steady-state number density of dust originating from the prograde Himalia family. The $x$-axis points toward the Sun, the $z$-axis is normal to the Jovian orbital plane, and the $y$-axis is completed by the right-hand rule. The red dashed line represents the orbit of Callisto, while the blue line denotes the orbital distance of the Himalia family. The result presented here is vertically averaged over $[-50,50]\,R_\mathrm{J}$, where most of the dust particles are located (see later in Fig.~\ref{fig:opticaldepth}). (b) Same as (a), but for the sum of dust from three retrograde satellite groups. The four colored lines from the inside out stand for the orbit of Callisto and three retrograde families of satellites, i.e.~Ananke, Carme and Pasiphae families, respectively.}
    \label{fig10}
\end{figure}

\begin{figure}
	\includegraphics[width=\columnwidth]{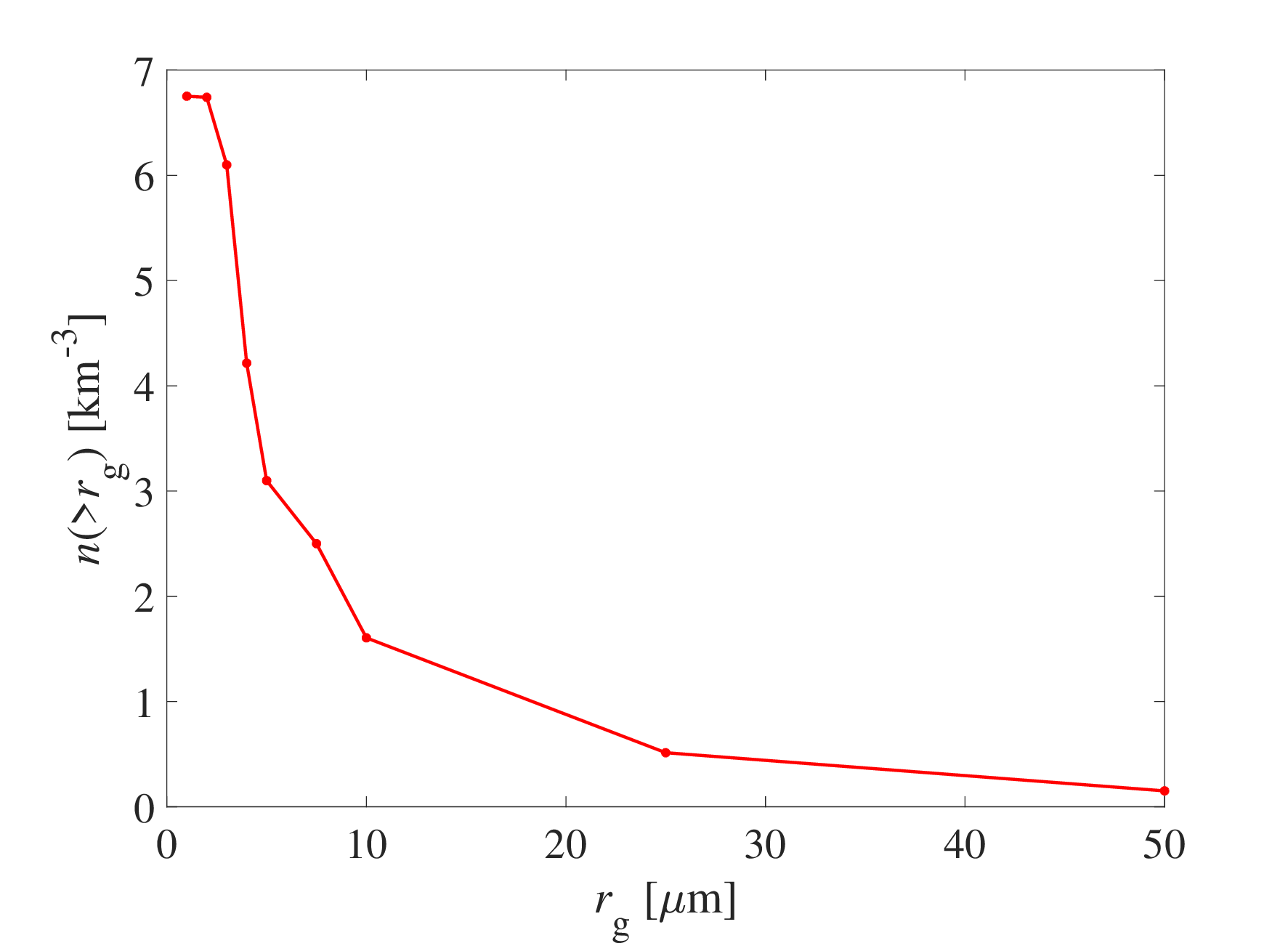}
    \caption{Cumulative dust number density of the torus formed by dust particles from Jupiter's irregular satellites, which is averaged over the radial range of $[30, 50]\,R_\mathrm{J}$, the vertical range of $[-50, 50]\,R_\mathrm{J}$, and the azimuthal angle range of [$0^{\circ}, 360^{\circ}$].}
    \label{fig11}
\end{figure}

% As shown in Fig.~\ref{fig10}, dust particles from irregular satellites form a torus with inner edge close to Callisto’s orbit due to the scavenging effect of Galilean satellites.
As shown in Fig.~\ref{fig10}, dust particles from irregular satellites form a torus in the radial range of about $[30, 50]\,R_\mathrm{J}$, with inner edge close to Callisto’s orbit due to the sweeping effect of Galilean satellites. The average number density of this torus is calculated (averaged over the radial range of $[30, 50]\,R_\mathrm{J}$, the vertical range of $[-50, 50]\,R_\mathrm{J}$, and the azimuthal angle range of [$0^{\circ}, 360^{\circ}$]), which is about $6.8\,\mathrm{km^{-3}}$ (see Table \ref{tab:ringcontribution}). The distribution of prograde satellite ejecta is shifted away from the Sun, while it is offset toward the Sun for dust from the retrograde satellites (see Fig.~\ref{fig10}).
%The reason for this difference is that the small particles, which are greatly affected by solar radiation pressure and solar gravity, dominate the torus.
The reason for this difference is explained as follows. As shown in Fig.~\ref{fig11}, the cumulative number density of the torus shows a rapid decrease between 2 and 25 $\mu$m, which indicates the predominance of particles in this size range.
% The radiation pressure and solar gravity strongly affect the orbital evolution of dust with size between 2 and 25 $\mu$m.
According to the same analytical theory applied in Section \ref{sec:final fate}, dust particles from the prograde Himalia family in the range of $[2, 25]\,{\mu}$m get the largest eccentricity when the solar angle equals to zero under the effect of solar radiation pressure and solar gravity; while for initial retrograde particles, the orbits become most eccentric when the solar angle is between $90^{\circ}$ and $180^{\circ}$ \cite[]{hamilton1996circumplanetary}.
As a consequence, prograde and retrograde satellite ejecta show different distribution trends relative to the Sun.
%The order of magnitude of the number density within $[30, 50]\,R_\mathrm{J}$ of dust from the prograde Himalia family is $10^{-8}\,\mathrm{m^{-3}}$, which is larger than that of the sum of dust from three retrograde satellite groups.

\begin{figure}
	\includegraphics[width=\columnwidth]{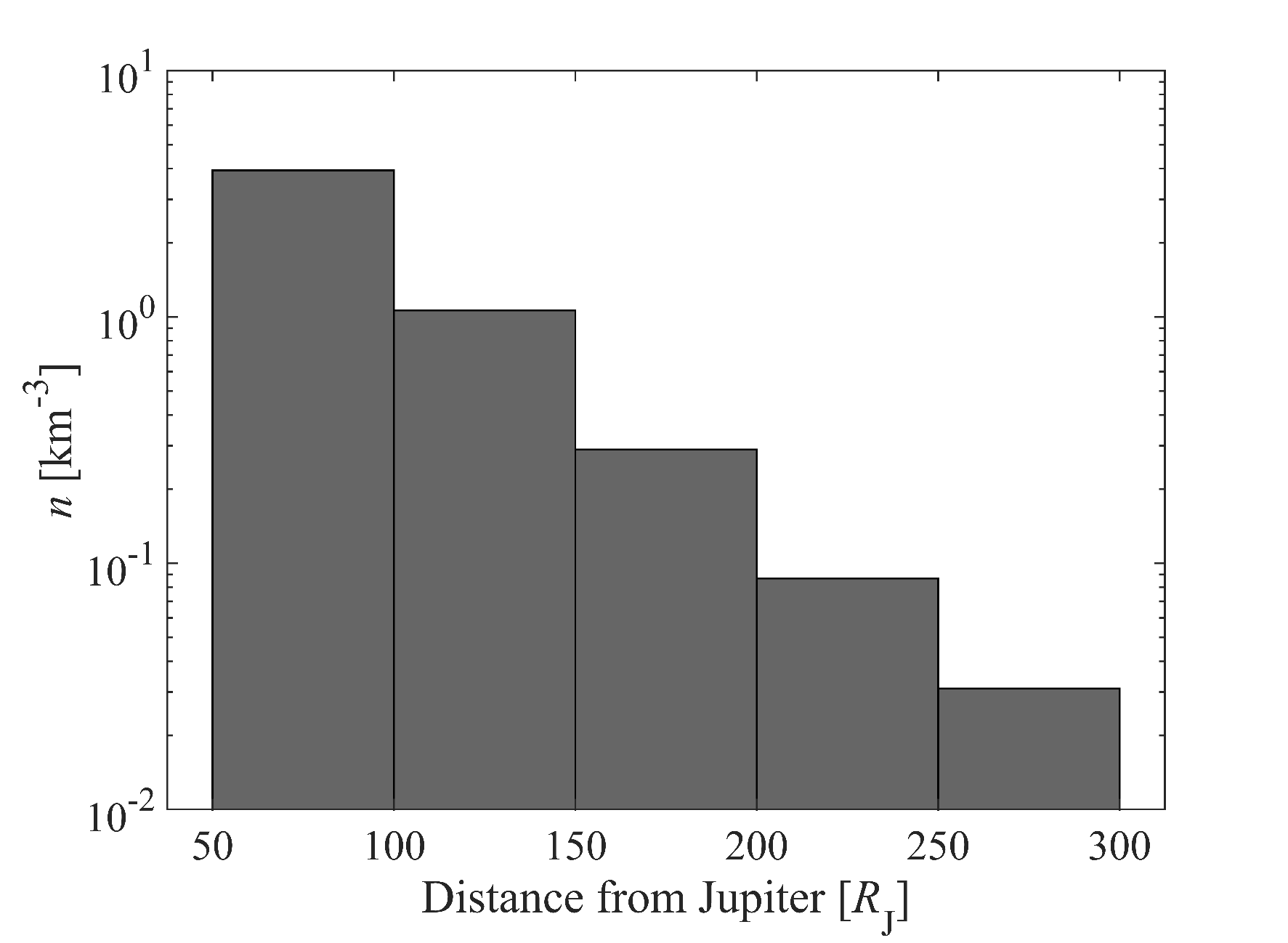}
    \caption{The radial number density profile between 50 and 300 $R_\mathrm{J}$ of dust ejected from Jupiter’s irregular satellites, where the number density is averaged over the vertical range of $[-50, 50]\,R_\mathrm{J}$ and the azimuthal angle range of [$0^{\circ}, 360^{\circ}$].}
    \label{fig:numberdensity}
\end{figure}

The average number density of dust from Jupiter's irregular satellites derived from our numerical simulation (Fig.~\ref{fig:numberdensity}) is lower than the number density in the outer region of the Jovian system estimated by \cite{krivov2002dust} based on the data obtained by the Dust Detection System (DDS) onboard the Galileo spacecraft. The result by \cite{krivov2002dust} has a nearly flat number density profile between 50 and 300 $R_\mathrm{J}$, which is about a level of $10\,\mathrm{km^{-3}}$. Our result shows that the number density between 50 and 100 $R_\mathrm{J}$ is about $4\,\mathrm{km^{-3}}$, and decreases as the distance from Jupiter increases. In fact, a declining radial profile of number density of dust from Jupiter's irregular satellites is reasonable since the inward migration caused by PR drag decelerates as the dust moves close to the planet, which means that dust particles spend longer time in the region closer to Jupiter. The difference in the number density profile may indicates that dust particles from other source bodies also contribute to the number density in the outer region of the Jovian system, or may simply attribute to the uncertainties of our modelling or the processing of the DDS data, or both.
%The small differences between the average number density of dust from Jupiter's irregular satellites and the number density derived from the data of DDS within this region may come from the error in the processing of the DDS's data and the uncertain parameters that used for calculating the production rate in this paper, including the bulk densities of parent satellite families and the mass flux and average velocity of the interplanetary dust particles.

%may indicates the other sources besides the irregular satellites of Jupiter of dust in this region.}

%(1) The data of DDS is limited, contaminated by interplanetary and interstellar dust and incomplete.

%\textcolor{violet}{Considering that the data of DDS is limited and incomplete transmission, the number density may be underestimated by \cite{krivov2002dust}.} \textcolor{violet}{In addition, the interplanetary dust flux $F_\mathrm{imp}^{\infty}=10^{-16}\,\mathrm{kg}\,\mathrm{m^{-2}}\,\mathrm{s^{-1}}$ that used for calculating the production rate in this paper is the maximum value within its error bars \cite[]{poppe2016improved}, which would cause us to overestimate the number density.}

%Besides, considering that the data of DDS is limited and incomplete transmission, the order of number density may be underestimated by \cite{krivov2002dust}.
%The most uncertainty of our result comes from the interplanetary dust environment model that used for calculating the production rate.

\subsection{Impact sites and accretion of dust on Callisto’s surface}
\label{sec:Callisto’s surface}
To further investigate the connection between the irregular satellites and the properties of the surfaces of Galilean satellites, the impact sites of dust are also recorded in our numerical simulations.

\begin{figure}
	\includegraphics[width=\columnwidth]{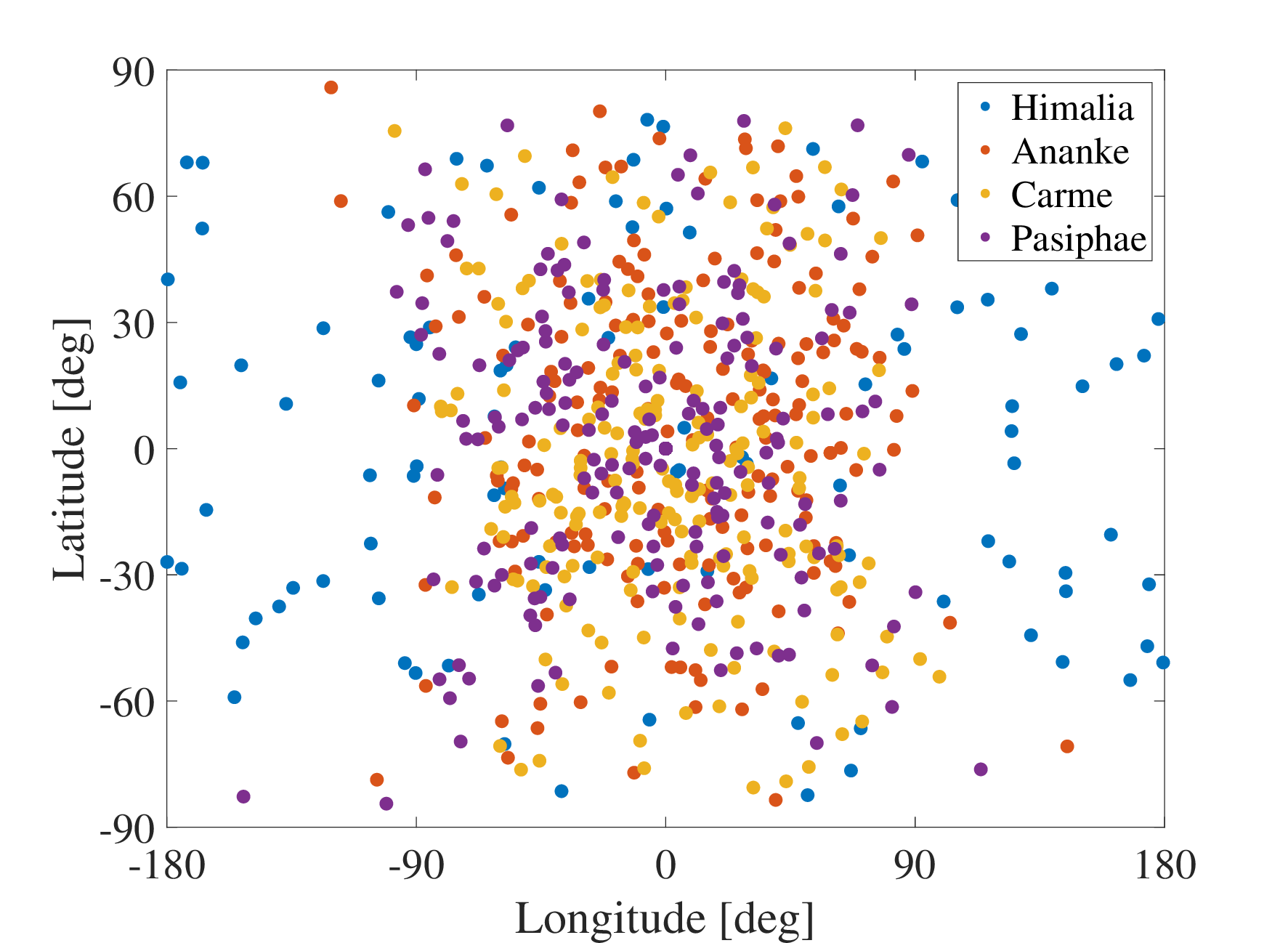}
    \caption{The impact sites of dust particles from different irregular satellite families on the surface of Callisto. The longitudes and latitudes of the impact sites are recorded in the satellite-centered frame $Ox_\mathrm{sat}y_\mathrm{sat}z_\mathrm{sat}$.%, where the $y$-axis always points to the planet, the $z$-axis is the spin axis of the satellite, and the $x$-axis is completed by right-hand rule and aligned with the direction of the moon’s motion.
    }
    \label{fig:impactsite}
\end{figure}

The orbits of the Galilean satellites are precessing during the numerical integration. 
%, with the initial condition adopted from \url{https://ssd.jpl.nasa.gov/sats/elem/}.
% The longitude and latitude of impact site in the satellite-centered frame that keeps the $y$-axis points to the planet are calculated, given that the Galilean satellites are tidally locked.
% The $z$-axis of the frame is the spin axis of the satellite, and the $x$-axis is completed by right-hand rule, which aligns with the direction of the moon’s motion.
Given that the Galilean satellites are tidally locked, a satellite-centered frame ($Ox_\mathrm{sat}y_\mathrm{sat}z_\mathrm{sat}$) that keeps the $y$-axis pointing to the planet is used for calculating the impact sites of dust.
The $z$-axis of the frame is along the spin axis of the satellite, and the $x$-axis is completed by right-hand rule and points to the direction of the moon’s motion.
% The impact sites of dust are recorded as the longitude which specifies the east–west position relative to the $x$-axis and the latitude which measures the north–south position relative to the $x-y$ plane.
The longitudes and the latitudes of the impact sites of dust are recorded in the $Ox_\mathrm{sat}y_\mathrm{sat}z_\mathrm{sat}$ frame.
% Considering that most of the dust are cleared by Callisto and that the characteristics of the surface of the other three moons are strongly affect by their more active geological activity \cite[]{lopes2014io,jia2018evidence,ligier2019surface} and the complicated environment as they are closer to the Jupiter \cite[]{bagenal2020space,paranicas2009europa,ligier2019surface},
Considering that most of the dust grains are cleared by Callisto in our simulation (see Section \ref{sec:final fate}) and that the characteristics of the surfaces of other three Galilean moons are strongly affected by other factors, including the more active geological activity \cite[]{lopes2014io,jia2018evidence,ligier2019surface} and the more complicated environment \cite[]{bagenal2020space,paranicas2009europa,ligier2019surface}, only the impact sites of dust on Callisto's surface are reported here.
As shown in Fig.~\ref{fig:impactsite}, the impact sites of dust ejected from the retrograde families of satellites are concentrated on the longitude between $-90^{\circ}$ and $90^{\circ}$, i.e., the leading hemisphere of Callisto; while dust from the prograde Himalia group are nearly uniformly distributed on the whole surface of Callisto. This difference of distribution comes from the different collision mechanisms between the dust from retrograde and prograde satellites (see Section \ref{sec:final fate}).

Previous studies have shown that the leading hemisphere of Callisto is darker than its trailing hemisphere \cite[]{bagenal2007jupiter}, and the compositions of these two hemispheres are different \cite[]{hibbitts2000distributions,cartwright2020evidence}.
% \textcolor{red}{and the distribution of \textcolor{violet}{$\mathrm{CO_2}$ and $\mathrm{HS_2}$} on Callisto's surface is hemispherically asymmetrical} \cite[]{hibbitts2000distributions,cartwright2020evidence}.
%some species, including $\mathrm{CO_2}$ and $\mathrm{HS_2}$, on Callisto's surface is hemispherically asymmetrical} \cite[]{hibbitts2000distributions,cartwright2020evidence}.
%and the composition of the two hemispheres exists difference \cite[]{hibbitts2000distributions,cartwright2020evidence}.
But the research about the composition of Jupiter’s irregular satellites' surface is limited \cite[]{brown20143,bhatt2017composition,grav2004near,grav2015neowise}, especially the retrograde moons, that we cannot directly assert from our result that the leading-trailing asymmetry of Callisto’s surface is contributed by dust from the irregular satellites.
%\textcolor{red}{Besides, if the material of dust will change during migration due to the sublimation or space environment is also unclear.}
Nevertheless, our result provides an important clue to the study of the hemispherical asymmetry of Callisto.

%\subsection{The accretion rate of dust on Callisto}
%\textcolor{violet}{Based on the calculation of the number density of dust from Jupiter's irregular satellites, we can estimate the number of these dust particles captured by four Galilean satellites per unit time, which is the further work of Section 4.3. With the same considerations, i.e., most of the dust grains are cleared by Callisto, as in Section 4.3, only the mass accretion rate, cross-sectional area influx and mass influx density of dust on Callisto are calculated here.}

The number of particles from Jupiter's irregular satellites captured by Callisto per unit time reads \citep{kortenkamp1998accretion}
\begin{equation}
    p = {\sigma}nv_\mathrm{o}.
\end{equation}
Here ${\sigma}$ is the effective capture cross-section of Callisto, $n$ is the number density of particles near Callisto, and $v_\mathrm{o}$ is the relative speed between dust and Callisto.
% $v_\mathrm{o} = \lvert \vec{v}_\mathrm{dust}-\vec{v}_\mathrm{Calliso} \rvert$ is the relative speed between dust and Callisto, where the \textcolor{cyan}{orbital} velocity of Callisto $v_\mathrm{Calliso}$ is about $8.2\,\mathrm{km\,s^{-1}}$.}

The effective capture cross-section of Callisto is calculated by considering the effect of gravitational focusing by the moon, which was given by \cite[]{opik1951collision}
\begin{equation}
    {\sigma}={\pi}R_\mathrm{Ca}^2\left(1+\frac{v_\mathrm{esc}^2}{v_\mathrm{o}^2}\right),
\end{equation}
where $R_\mathrm{Ca}=2410.3\,\mathrm{km}$ and $v_\mathrm{esc}=2.44\,\mathrm{km\,s^{-1}}$ are the radius and escape velocity of Callisto, respectively.

Considering the contributions from dust particles of all grain sizes, the mass accretion rate $M_\mathrm{accret\_rate}$ and the cross-sectional area influx $A_\mathrm{accret\_rate}$ of dust particles on Callisto are calculated by
% Considering that $p$ is grain-size-dependent, 
%\textcolor{cyan}{ is calculated by integrating $m\mathrm{_g\,d}p$ over all grain sizes, which reads}
\begin{equation}
M_\mathrm{accret\_rate}=\int_{r_\mathrm{min}}^{r_\mathrm{max}} m_\mathrm{g}\,\mathrm{d}p,
\end{equation}
\begin{equation}
A_\mathrm{accret\_rate}=\int_{r_\mathrm{min}}^{r_\mathrm{max}} A_\mathrm{g}\,\mathrm{d}p.
\end{equation}
Besides, the mass influx density $F_\mathrm{accret\_rate}$ that quantifies the mass of dust accreted by Callisto per unit surface area and per unit time is roughly estimated. For dust particles from the prograde Himalia family, $F_\mathrm{accret\_rate} = M_\mathrm{accret\_rate}/4\pi R_\mathrm{Ca}^2$ because these particles are nearly uniformly distributed on the whole surface of Callisto; while for dust particles from the three retrograde families, $F_\mathrm{accret\_rate} = M_\mathrm{accret\_rate}/2\pi R_\mathrm{Ca}^2$ because these particles are only distributed on the leading hemisphere of Callisto (see Figure \ref{fig:impactsite}).
% by dividing $M_\mathrm{accret\_rate}$ by Callisto's surface area. 
% Note that it should be $M_\mathrm{accret\_rate}$ divided by half the surface area of Callisto for dust from the three retrograde satellite families, by taking into account the hemispherical asymmetry of the impact sites of these particles (Section 4.3).

\begin{figure}
	\includegraphics[width=\columnwidth]{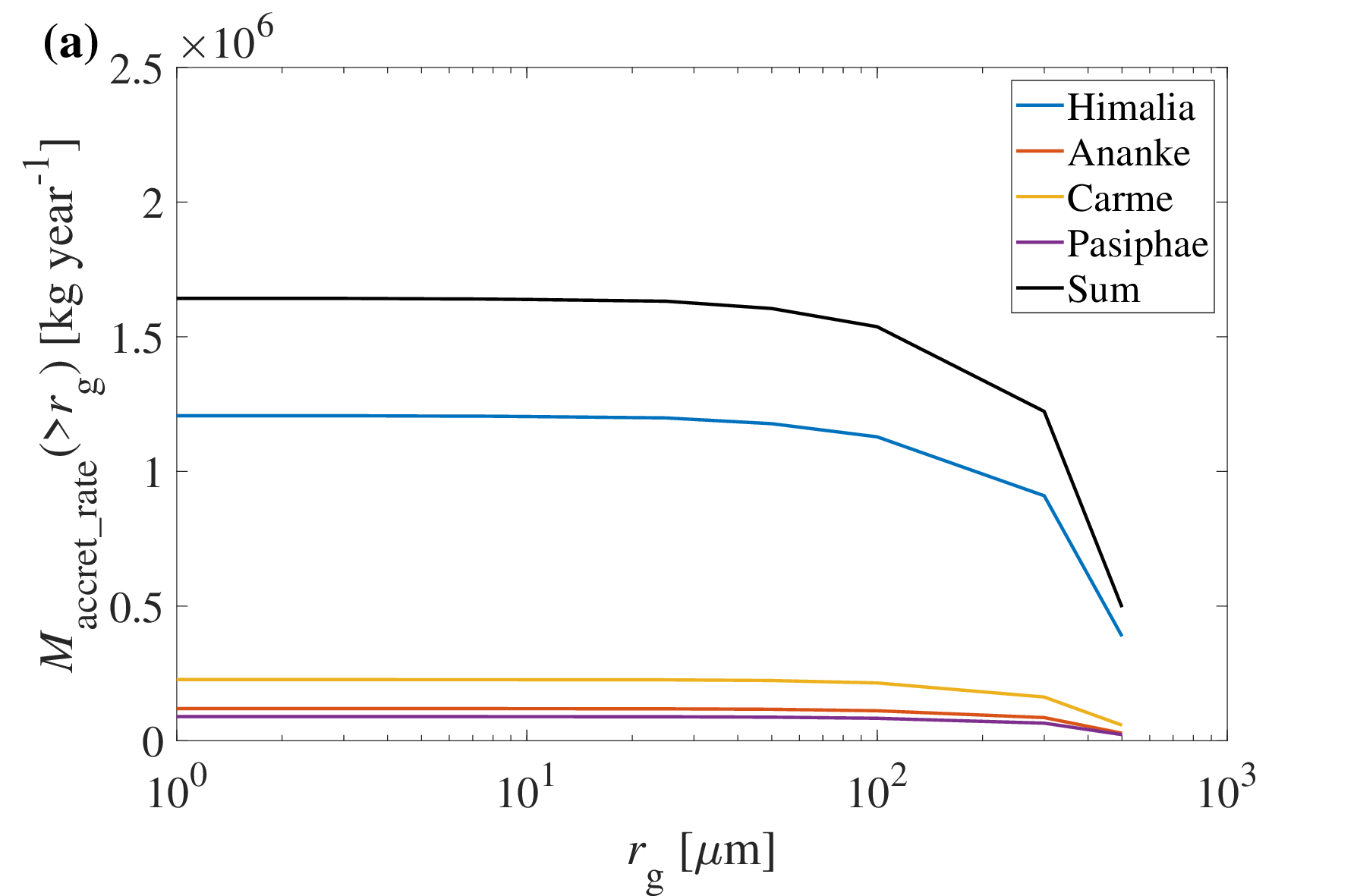}
        \includegraphics[width=\columnwidth]{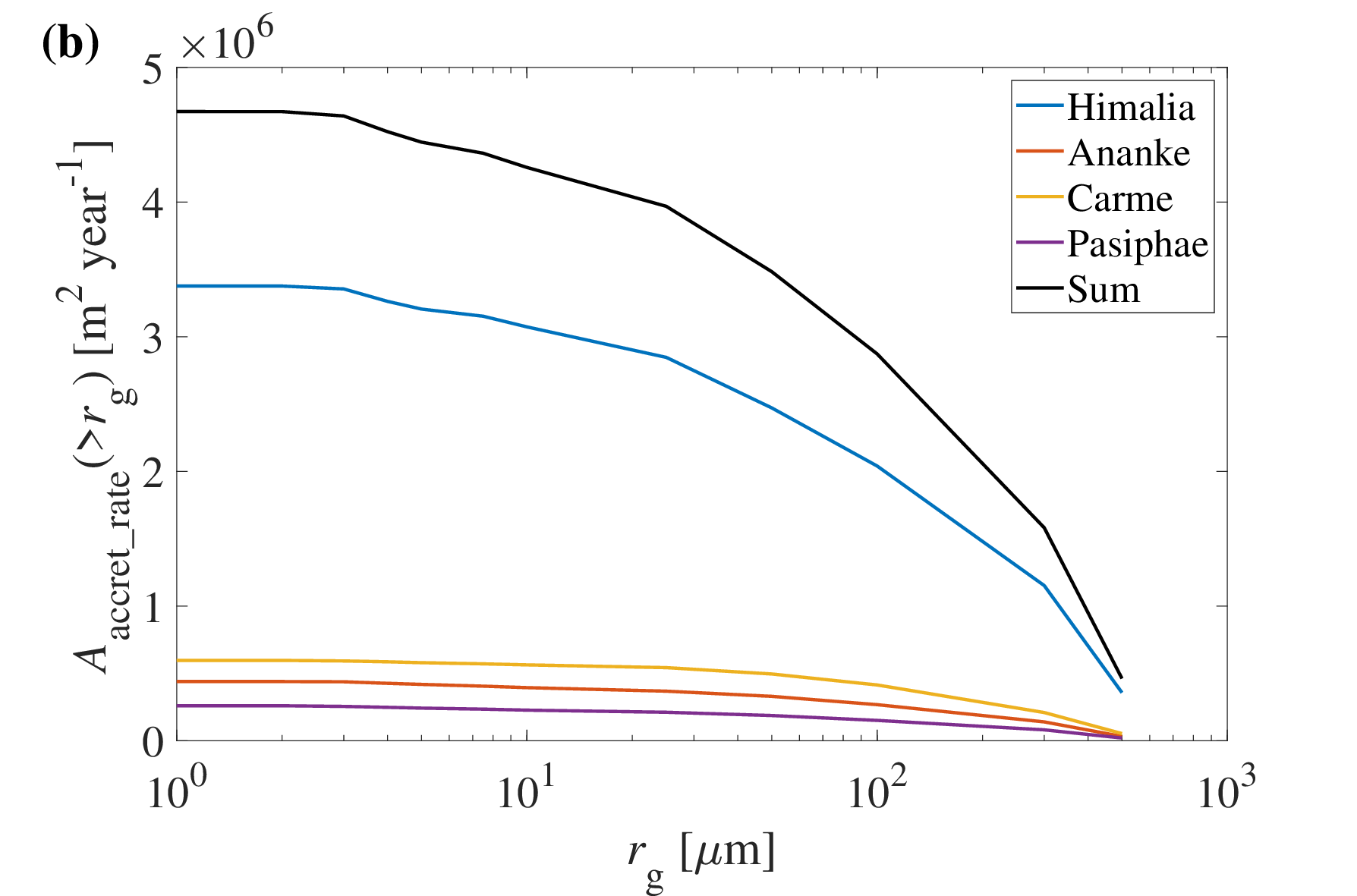}
        \includegraphics[width=\columnwidth]{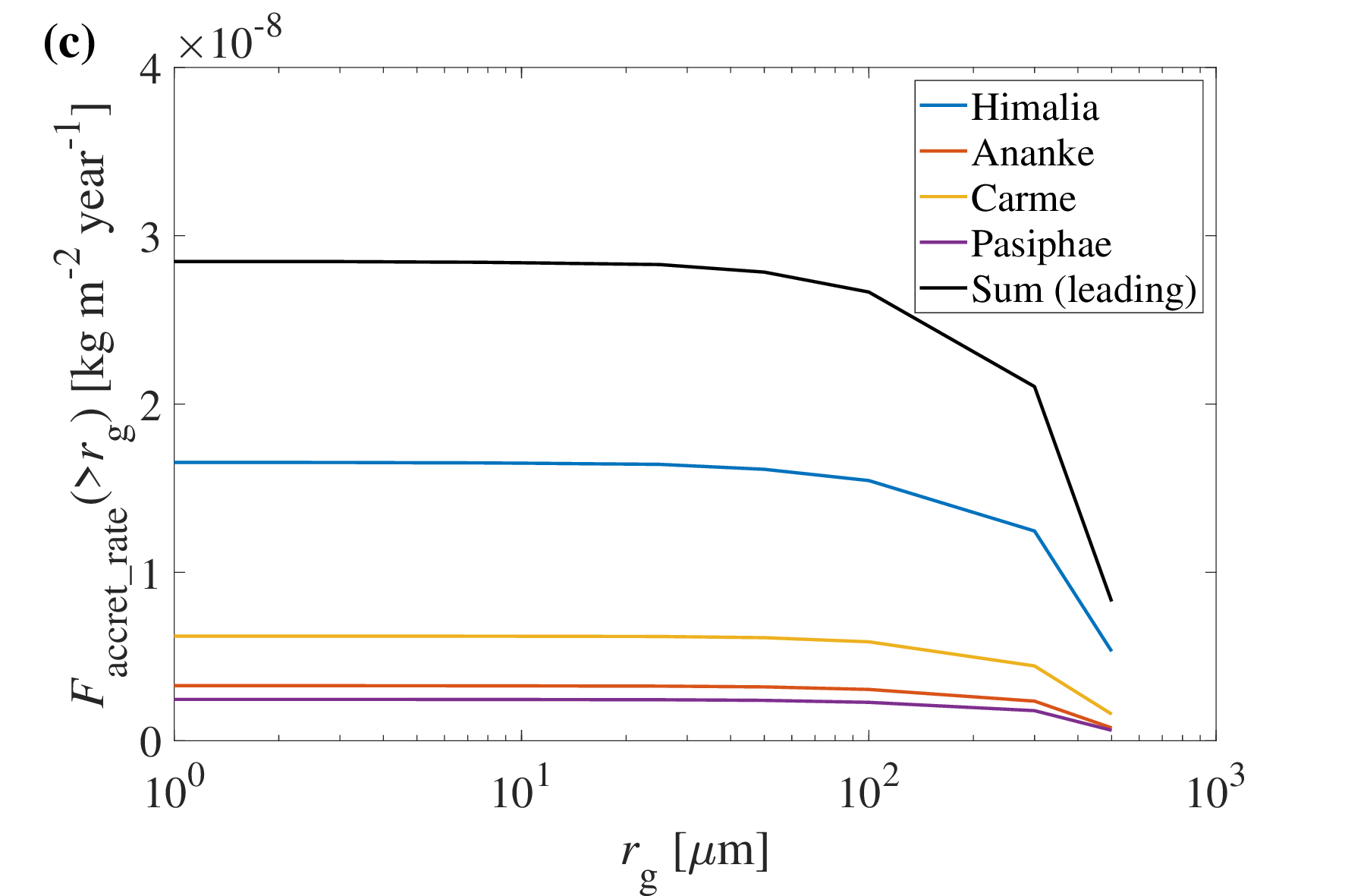}
    \caption{Cumulative distributions of mass accretion rate (a), cross-sectional area influx (b) and mass influx density (c) of dust from different irregular satellite families on Callisto and their sum. Note that the black line in (c) denotes the total $F_\mathrm{accret\_rate}$ on the leading hemisphere of Callisto, while on Callisto's trailing hemisphere the total $F_\mathrm{accret\_rate}$ is approximately equal to the one contributed by dust from the prograde Hiamlia family.}
    \label{fig:accretion}
\end{figure}

The cumulative distributions of the mass accretion rate $M_\mathrm{accret\_rate}$, cross-sectional area influx $A_\mathrm{accret\_rate}$ and mass influx density $F_\mathrm{accret\_rate}$ of dust from Jupiter's irregular satellites on Callisto are shown in Fig.~\ref{fig:accretion}.
The total $M_\mathrm{accret\_rate}$ and $A_\mathrm{accret\_rate}$ on Callisto are about $1.6\times 10^6\,\mathrm{kg\,year^{-1}}$ and $4.7\times 10^6\,\mathrm{m^{2}\,year^{-1}}$, respectively. On the leading hemisphere of Callisto, $F_\mathrm{accret\_rate}$ is about $2.8\times 10^{-8}\,\mathrm{kg\,m^{-2}\,year^{-1}}$, and on the trailing hemisphere of Callisto, it is about $1.7\times 10^{-8}\,\mathrm{kg\,m^{-2}\,year^{-1}}$. It can be easily seen that the values of $M_\mathrm{accret\_rate}$, $A_\mathrm{accret\_rate}$ and $F_\mathrm{accret\_rate}$ of dust from the prograde Himalia family on Callisto are much higher than those from the three retrograde families of satellites. For each irregular satellite family, large dust particles contribute more to $M_\mathrm{accret\_rate}$, $A_\mathrm{accret\_rate}$ and $F_\mathrm{accret\_rate}$. Especially, for $M_\mathrm{accret\_rate}$ and $F_\mathrm{accret\_rate}$, particles larger than $100\,{\mu}$m are dominant.

%The mass accretion rates of dust from different irregular satellite families on Callisto are estimated and listed in Table \ref{tab:accretion}. It is seen that the mass of dust from the prograde Himalia family that impact Callisto per unit time are larger than those from the three retrograde families of satellites. The total mass of dust from Jupiter's irregular satellites accreted by Callisto per unit time is about a level of $10^{-8}\,\mathrm{kg\,s^{-1}}$.

%\begin{table}
%\renewcommand{\arraystretch}{1.3}
%	\centering
	%\caption{This is an example table. Captions appear above each table.
	%Remember to define the quantities, symbols and units used.}
 %       \caption{\textcolor{cyan}{The mass accretion rate $(M_\mathrm{accret\_rate})$, cross-sectional area influx $(A_\mathrm{accret\_rate})$ and mass influx density $(F_\mathrm{accret\_rate})$ of dust from Jupiter's irregular satellites on Callisto, including the values of dust from four satellite families and their sum.}}
        
  %      \label{tab:accretion}
   %     \resizebox{\linewidth}{!}{
	%\begin{tabular}{lccccr} % four columns, alignment for each
	%	\hline
	%	& Himalia & Ananke & Carme & Pasiphae & Sum\\
	%	\hline
         %   $M_\mathrm{accret\_rate}\,[\mathrm{kg\,year^{-1}}]$ & 2.3E-1 & 4.4E-2 & 3.2E-2 & 8.2E-2 & 3.9E-1\\
          %  $A_\mathrm{accret\_rate}\,[\mathrm{m^2\,year^{-1}}]$ & 5.1E1 & 1.0E1 & 9.0E0 & 1.8E1 & 8.8E1\\
           % $F_\mathrm{accret\_rate}\,[\mathrm{kg\,m^{-2}\,year^{-1}}]$ & 1.3E-8 & 2.4E-9 & 1.7E-9 & 4.5E-9 & 2.2E-8\\
		%\hline
	%\end{tabular}}
%\end{table}

\subsection{If there is a ring?}
\label{sec:ring?}
\begin{figure}
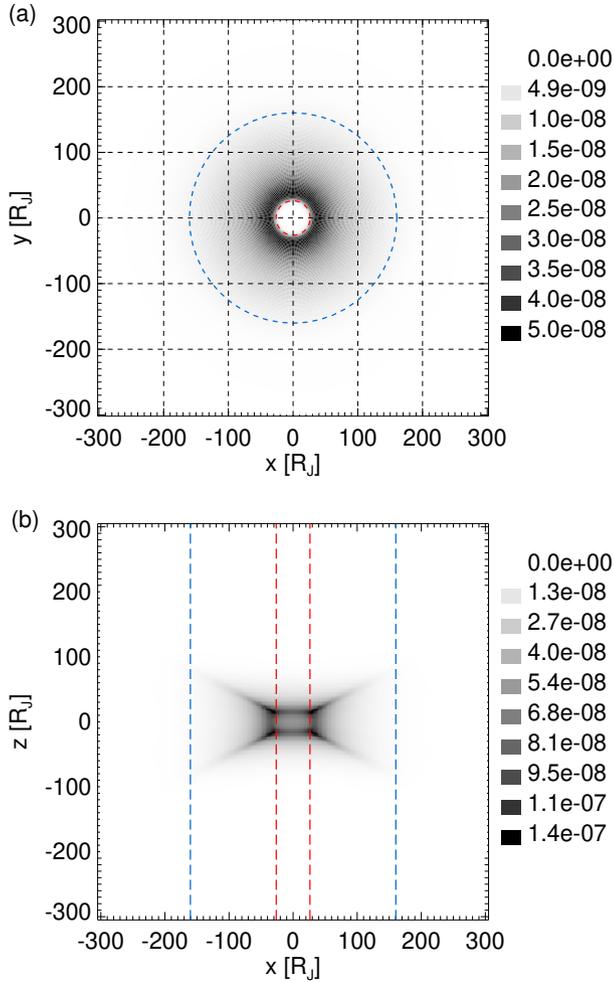

	\includegraphics[width=\columnwidth]{Fig13a.eps}
        \includegraphics[width=\columnwidth]{Fig13b.eps}
    \caption{(a) Normal geometric optical depth of the torus formed by the sum of dust from four irregular satellite families. The frame used here is $Ox_\mathrm{sun}y_\mathrm{sun}z_\mathrm{sun}$. The red and blue dashed lines denote the orbits of Callisto and the Himalia family, respectively. (b) Geometric optical depth when viewed toward the positive direction of the $y$-axis. The inner two red lines denote Callisto’s orbital distance, and the two blue lines denote the orbital distance of the Himalia family.}
    \label{fig:opticaldepth}
\end{figure}

The geometric optical depth of the torus outside the orbit of Callisto formed by the grains from irregular satellites is obtained by dividing the total cross-sectional area of all particles that are present in the grid by the surface area $A(i_\mathrm{cell},j_\mathrm{cell},k_\mathrm{cell})$ of the cell in the line-of-sight, which reads
\begin{equation}
    \resizebox{\columnwidth}{!}{$
        \begin{split}        
\tau(i_\mathrm{cell},j_\mathrm{cell},k_\mathrm{cell})=\int_{r_\mathrm{min}}^{r_\mathrm{max}}\frac{\tilde{n}(i_\mathrm{cell},j_\mathrm{cell},k_\mathrm{cell}; r_\mathrm{g}){\pi}r_\mathrm{g}^2}{A(i_\mathrm{cell},j_\mathrm{cell},k_\mathrm{cell})n_\mathrm{start}}{\Delta}t N(r_\mathrm{g})\mathrm{d}r_\mathrm{g}.
        \end{split}
    $}
\end{equation}
The simulation results of the geometric optical depth are also shown in $Ox_\mathrm{sun}y_\mathrm{sun}z_\mathrm{sun}$ (Fig.~\ref{fig:opticaldepth}). The normal geometric optical depth of the torus decreases as the distance from Jupiter increases, the average value of which over the brightest region $[30, 50]\,R_\mathrm{J}$ is about $4.5\times 10^{-8}$ (Table \ref{tab:ringcontribution}). When viewed toward the positive direction of the $y$-axis, the average edge-on geometric optical depth in the region of $[30, 50]\,R_\mathrm{J}$ is about $6.5\times 10^{-8}$ (Table \ref{tab:ringcontribution}), which is comparable with Phoebe ring's edge-on optical depth \cite[]{verbiscer2009saturn}. The upper and lower edges of ring ansae are the brightest in the edge-on view, which is similar to Jupiter’s gossamer rings \cite[]{burns1999formation, showalter2008properties}.

%$10^{-8}$, which is comparable with that of the Phoebe ring \cite[]{verbiscer2009saturn}.
%When viewed toward the positive direction of the $y$-axis, the optical depth in the region of $[30, 50]\,R_\mathrm{J}$ is up to the order of $10^{-8}$, which is comparable with Phoebe ring's normal optical depth \cite[]{verbiscer2009saturn}, and the upper and lower edges of ring ansae are the brightest, which is similar to the Jupiter’s gossamer rings \cite[]{burns1999formation}.
This ring outside Callisto’s orbit formed by the dust from Jupiter's irregular satellites is probably observable. Because its geometric optical depth is on the order of observable magnitude, and its radial profile of number density is similar to that of the observed Phoebe ring \cite[]{tamayo2016radial}, which also declines from the outermost regular moon Iapetus to the irregular satellites region.

% Besides, the production rate of dust of the outer moons of Jupiter is higher than that of Saturn's, as the flux of interplanetary micrometeoroids is larger.
% However, there is no any direct or indirect evidence for a dust ring outside Callisto’s orbit yet.
% Only a prediction that dust from collisions between irregular moons can reach detectable levels around the giant planet was made by \cite{kennedy2011collisional}.
% Nevertheless, the existence of the Phoebe ring which \textcolor{red}{that} has an inner edge close to Iapetus and a similar radial profile with the torus described above \cite[see][]{tamayo2016radial} leads us to believe that there should be an observable ring outside Callisto’s orbit consisted of dust ejected from the Jupiter’s irregular satellites.
%Because the dust production rates is higher of the outer moons of Jupiter, as the flux of interplanetary micrometeoroids is larger and the surface area of Himalia and Phoebes’ are comparable.
%In addition, the distance for dust from Jupiter's irregular moons to transport inward is shorter than dust from Phoebe (Phoebe’s mean semimajor axis is 222 Saturnian radii, and the value of Jupiter’s irregular moons are shown in Fig.~\ref{fig1}),
%which indicates the less chance for dust destruction by sputtering, sublimation or mutual collision.

The contributions of dust from different irregular satellite families to the ring are calculated (see Table \ref{tab:ringcontribution}). It is found that dust particles from the prograde Himalia family are dominant in the ring, and their contributions to the ring (including the ring's average number density, normal and edge-on geometric optical depths) are about one order of magnitude higher than the sum of those of the three retrograte satellite families.

\begin{table}
\renewcommand{\arraystretch}{1.2}
	\centering
	%\caption{This is an example table. Captions appear above each table.
	%Remember to define the quantities, symbols and units used.}
        \caption{The average values of number density $(n)$, normal geometric optical depth $(\tau_\mathrm{normal})$ and edge-on geometric optical depth $(\tau_\mathrm{edge-on})$ for the dust torus outside Callisto’s orbit. The contributions of dust from different irregular satellite families and their sum are listed.}
        \label{tab:ringcontribution}
        \resizebox{\linewidth}{!}{
	\begin{tabular}{lccccr} % four columns, alignment for each
		\hline
		&Himalia & Ananke & Carme & Pasiphae & Sum\\
		\hline
            $n\,[\mathrm{{km}^{-3}}]$ & 5.7E0 & 4.1E-1 & 2.7E-1 & 4.2E-1 & 6.8E0\\
            $\tau_\mathrm{normal}$ & 4.0E-8 & 2.3E-9 & 1.5E-9  & 1.6E-9 & 4.5E-8\\
            $\tau_\mathrm{edge-on}$ & 5.6E-8 & 3.5E-9 & 2.9E-9 & 2.8E-9 & 6.5E-8\\
		\hline
	\end{tabular}}
\end{table}
% \textcolor{cyan}{These quantities are averaged over the radial range of $[30, 50]\,R_\mathrm{J}$, the vertical range of $[-50, 50]\,R_\mathrm{J}$, and the azimuthal angle range of [$0^{\circ}, 360^{\circ}$].}

\section{Summary}
In this paper, we analyze the "life" of dust particles ejected from the irregular satellites of Jupiter by the impact of interplanetary micrometeoroids, including their birth, dynamical evolution, sink, and steady-state spatial distribution.
%The production rate of dust from four satellite families is calculated by the formula suggested by \cite{krivov2003impact}, and the initial size distribution is obtained by being normalized to the production rate.
The production rates of dust from four satellite families are calculated, and the initial size distribution of dust is obtained.
The high-accuracy numerical integration of the dynamical equation of dust is performed.
%The perturbation forces, including the solar radiation pressure, gravity due to the Sun and Galilean satellites, Jovian oblateness and Poynting-Robertson drag, are considered.
The trajectories of dust, with 12 grain sizes in the range $[1, 500]\,{\mu}$m and 200 different initial orbits for each size, are simulated until they hit a sink (impact the Galilean satellites, Jupiter or escape from the Jovian system).
% 尺寸分布的范围要说一下% 已修改
By integrating the trajectories of dust over their initial size distribution, the spatial number density and the geometric optical depth are estimated. Combined with the analytical theory, our results are summarized as follows,

%(1) Jupiter’s prograde irregular satellite family, Himalia family, produces more particles per unit second compared to \textcolor{red}{than} retrograde families, because of the stronger gravitational focusing effect upon the impactor in Himalia region and the larger cross-section suffered to impact of Himalia family.
(1) The mass production rate of Jupiter’s prograde irregular satellite family, the Himalia family, is higher than those of the retrograde families, because of the stronger effect of gravitational focusing by Jupiter on the impactor in the Himalia region and the larger cross-section suffered to impact of the Himalia family.

(2) The average dynamical lifetimes of small particles rely on the strengths of solar radiation pressure and solar gravity. Especially, particles smaller than 2 $\mu$m have a short life because of the strong influence induced by solar radiation pressure. For dust larger than a few micrometers, its dynamical lifetime is approximately equal to the inward migration time due to the PR drag.

%(3) Large dust particles ($r_\mathrm{g}>5\,{\mu}$m) from the Pasiphae family with initial eccentricity larger than 0.4 and initial solar angle close to $0^{\circ}$ or $180^{\circ}$ get into the highest eccentric orbit under the perturbation of solar gravity, which leads to collision with Jupiter or escape from the Jovian system.
(3) The orbital eccentricities of large dust particles ($>5\,{\mu}$m) from the Pasiphae family with initial eccentricities larger than 0.4 and initial solar angles close to $0^{\circ}$ or $180^{\circ}$ get close to 1 under the perturbation of solar gravity, which leads to collide with Jupiter or escape from the Jovian system.

(4) As the dust migrates to the regions of Galilean satellites, dust particles from the prograde Himalia family become more unstable than that from the retrograde groups due to the gravitational perturbations of Galilean satellites. As a result, there are still approximately 20$\%$ initial prograde particles that impact Jupiter or escape from the Jovian system when the grain size is large, and the opportunities to hit the four Galilean moons for dust particles from the prograde satellites are about 1$\%$, 10$\%$, 20$\%$ and 50$\%$ respectively; while almost all of dust particles from the retrograde satellites collide with Callisto.

(5) The steady-state number density of dust from Jupiter's irregular satellites in the Jovian system is estimated.
Dust particles originating from Jupiter's irregular moons gather into a torus outside the orbit of Callisto.
The average number density of the torus is about $6.8\,\mathrm{km^{-3}}$. The appearance of torus is shifted away from the Sun for particles from the prograde irregular moons; while it is offset toward the Sun for retrograde satellite ejecta.
%This is because the dust in the size range of $[2, 25]\mu$m is dominant in the torus.
%This is because the dust in the size range of $[2, 25]\mu$m is dominant in the torus, which the orbits of prograde particles in this size range are elongated get the largest eccentricities when the solar angle equals to zero under the disturbance of radiation pressure and solar gravity; while the orbits of retrograde grains become most eccentric when the solar angle is between $90^{\circ}$ and $180^{\circ}$.

(6) The impact sites of dust from the prograde family (Himalia family) are nearly uniformly distributed on the whole surface of Callisto; while those of dust from the three retrograde families (Ananke, Carme, and Pasiphae families) are centered on the leading hemisphere of Callisto. The distribution of the impact sites of dust from retrograde irregular satellites provides an important clue for understanding the leading-trailing asymmetry of Callisto’s surface.
% The impact sites of dust from retrograde satellite families are centered on the leading hemisphere of Callisto, which provides an important clue for understanding the leading-trailing asymmetry of Callisto’s surface.

(7) The mass accretion rate, cross-sectional area influx and mass influx density of dust from Jupiter's irregular satellites on Callisto are estimated. The total mass accretion rate of dust on Callisto is about $1.6\times 10^6\,\mathrm{kg\,year^{-1}}$, and the total cross-sectional area influx on Callisto is about $4.7\times 10^6\,\mathrm{m^{2}\,year^{-1}}$. The total mass influx density are about $2.8\times 10^{-8}\,\mathrm{kg\,m^{-2}\,year^{-1}}$ and $1.7\times 10^{-8}\,\mathrm{kg\,m^{-2}\,year^{-1}}$ on Callisto's leading hemisphere and trailing hemisphere, respectively. Compared to Jupiter's retrograde families, dust particles from the prograde Himalia family contribute more to the total values of the mass accretion rate, cross-sectional area influx and mass influx density on Callisto. For each irregular satellite family, large dust particles contribute more to the mass accretion rate, cross-sectional area influx and mass influx density on Callisto.

(8) There is probably an observable ring outside Callisto’s orbit which consists of the dust ejected from Jupiter’s irregular satellites.
The ring is brightest in the radial range of about $[30, 50]\,R_\mathrm{J}$ and dominated by dust particles in the size range of $[2, 25]\,{\mu}$m.
The average normal geometric optical depth of the ring is about $4.5\times 10^{-8}$, and the average edge-on geometric optical depth is about $6.5\times 10^{-8}$.
% \textcolor{violet}{, which is comparable with Phoebe ring's light-of-sight optical depth \cite[]{verbiscer2009saturn}.}
% The configuration of the ring ansae is similar to the Jupiter’s gossamer rings. 
The upper and lower edges of ring ansae are the brightest in the edge-on view, which is similar to that of the Jupiter’s gossamer rings. Dust particles from the prograde Himalia family are dominant in the ring, and their contributions to the ring (including the ring's average number density, normal and edge-on geometric optical depths) are much larger than the sum of those of the three retrograte satellite families.

\section*{Acknowledgements}

%The Acknowledgements section is not numbered. Here you can thank helpful
%colleagues, acknowledge funding agencies, telescopes and facilities used etc.
This work was supported by the National Natural Science Foundation of China (No.~12002397, 12311530055, and 62388101), the National Key R\&D Program of China (No.~2020YFC2201202 and 2020YFC2201101), and by the Shenzhen Science and Technology Program (Grant No.~ZDSYS20210623091808026). 

%%%%%%%%%%%%%%%%%%%%%%%%%%%%%%%%%%%%%%%%%%%%%%%%%%
\section*{Data Availability}

%The inclusion of a Data Availability Statement is a requirement for articles published in MNRAS. Data Availability Statements provide a standardised format for readers to understand the availability of data underlying the research results described in the article. The statement may refer to original data generated in the course of the study or to third-party data analysed in the article. The statement should describe and provide means of access, where possible, by linking to the data or providing the required accession numbers for the relevant databases or DOIs.
The basic data of this work will be shared on reasonable request to the corresponding author.



\bibliographystyle{mnras}
\bibliography{example} % if your bibtex file is called example.bib

%\begin{thebibliography}{99}
%\end{thebibliography}

%%%%%%%%%%%%%%%%%%%%%%%%%%%%%%%%%%%%%%%%%%%%%%%%%%

%%%%%%%%%%%%%%%%% APPENDICES %%%%%%%%%%%%%%%%%%%%%

% \appendix

% \section{Some extra material}

% If you want to present additional material which would interrupt the flow of the main paper,
% it can be placed in an Appendix which appears after the list of references.

%%%%%%%%%%%%%%%%%%%%%%%%%%%%%%%%%%%%%%%%%%%%%%%%%%

% Don't change these lines
\bsp	% typesetting comment
\label{lastpage}
\end{document}